%% file: paper_part_I_noiseless.tex
\documentclass[12pt]{article}

\usepackage{graphicx}
\usepackage{amsfonts}
\usepackage{amssymb}
\usepackage{amsthm}
\usepackage{amsmath}
\usepackage{graphicx}
\usepackage{epsfig}
\usepackage{color}

\theoremstyle{plain}
\newtheorem{theorem}{Theorem} 
\newtheorem{lemma}[theorem]{Lemma}
\newtheorem{proposition}[theorem]{Proposition}
\newtheorem{corollary}[theorem]{Corollary}

\theoremstyle{definition}
\newtheorem{definition}[theorem]{Definition}

\theoremstyle{remark}

\definecolor{red}{named}{Red}
\definecolor{blue}{named}{BlueViolet}
\definecolor{green}{named}{LimeGreen}

\newcommand{\calV}{\mathcal{V}}
\newcommand{\calF}{\mathcal{F}}
\newcommand{\calS}{\mathcal{S}}
\newcommand{\calM}{\mathcal{M}}
\newcommand{\calI}{\mathcal{I}}
\newcommand{\calRs}{\mathcal{R}_{\square}}

\newcommand{\Z}{\mathbb{Z}}
\newcommand{\R}{\mathbb{R}}
\newcommand{\N}{\mathbb{N}}

\newcommand{\Ceil[1]}{\ensuremath{\left\lceil#1\right\rceil}}
\newcommand{\Floor[1]}{\ensuremath{\left\lfloor#1\right\rfloor}}

\newcommand{\defined}{\stackrel{\triangle}{=}}

\begin{document}
\title{Scanning and Sequential Decision Making for Multi-Dimensional Data - Part I: the Noiseless Case\thanks{The material in this
paper was presented in part at the IEEE International Symposium on Information Theory, Seattle, Washington, United States, July 2006, and the Electricity 2006 convention, Eilat, Israel, November 2006.}}
\author{Asaf Cohen\thanks{Asaf Cohen and Neri Merhav are with the Department of the Electrical Engineering, Technion - I.I.T., Haifa 32000, Israel. E-mails: \{soofsoof@tx,merhav@ee\}.technion.ac.il.}, Neri Merhav$^\dagger$ and Tsachy Weissman\thanks{Tsachy Weissman is with the Department of Electrical Engineering, Stanford University, Stanford, CA 94305, USA. E-mail: tsachy@stanford.edu.}}
\maketitle
\date
\input{abs.tex}
\input{introduction.tex}
\input{formulation.tex}
\input{universal.tex}
\input{sensitivity.tex}
\input{conclusion.tex}

\input{appendix.tex}

\bibliographystyle{../latex/IEEEbib}
\bibliography{../latex/prediction_and_coding}
\end{document}

%% file: abs.tex
\abstract{We investigate the problem of scanning and
prediction (``scandiction", for short) of multidimensional data arrays. This
problem arises in several aspects of image and video processing, such as
predictive coding, for example, where an image is compressed by coding the error sequence
resulting from scandicting it. Thus, it is
natural to ask what is the optimal method to scan and predict a given
image, what is the resulting minimum prediction loss, and whether there exist
specific scandiction schemes which are universal in some
sense.

Specifically, we investigate the following problems: First,
modeling the data array as a random field, we wish to examine whether there exists a scandiction scheme which is independent of the field's distribution,
yet asymptotically achieves the same performance as if this distribution
was known. This question is answered in the affirmative for the set of all spatially stationary random fields and under mild conditions on the loss function. We then discuss the scenario where a non-optimal scanning order is used, yet accompanied by an optimal predictor, and derive bounds on the excess loss compared to optimal scanning and prediction.

This paper is the first part of a two-part paper on sequential decision making for multi-dimensional data. It deals with clean, noiseless data arrays. The second part deals with noisy data arrays, namely, with the case where the decision maker observes only a noisy version of the data, yet it is judged with respect to the original, clean data.  

\vspace{0.3cm}\emph{Index Terms}-Universal scanning, Scandiction, Sequential decision making, Multi-dimensional data, Random Field, Individual image.

%% file: introduction.tex
\section{Introduction}\label{sec. intro.}Consider the problem of sequentially scanning and predicting a multidimensional data array, while minimizing a given loss function. Particularly, at each time instant $t$, $1 \leq t \leq |B|$, where $|B|$ is the number of sites (``pixels") in the data array, the scandictor chooses a site to be visited, denoted by $\Psi_t$, and gives a prediction, $F_t$, for the value at that site. Both $\Psi_t$ and $F_t$ may depend of the previously observed values - the values at sites $\Psi_1$ to $\Psi_{t-1}$. It then observes the true value, $x_{\Psi_t}$, suffers a loss $l(x_{\Psi_t},F_t)$, and so on. The goal is to minimize the cumulative loss after scandicting the entire data array.

The problem of sequentially predicting the next outcome of a one-dimensional sequence (or any data array with a fixed, predefined, order), $x_t$, based on the previously observed outcomes, $x_1,x_2,\ldots,x_{t-1},$ is well studied. The problem of prediction in multidimensional data arrays (or when reordering of the data is allowed), however, has received far less attention. Apart from the on-line strategies for the sequential prediction of the data, the fundamental problem of \emph{scanning} it should be considered. We refer to the former problem as the \emph{prediction} problem, where no reordering of the data is allowed, and to the latter as the \emph{scandiction} problem. 

The scandiction problem mainly arises in image compression, where various methods of predictive coding are used (e.g., \cite{Weinberger_Seroussi_Sapiro96}). In this case, the encoder may be given the freedom to choose the actual \emph{path} over which it traverses the image, and thus it is natural to ask which path is optimal in the sense of minimal cumulative prediction loss (which may result in maximal
compression). The scanning problem also arises in other areas of image processing, such as one-dimensional wavelet \cite{Lamarque_Robert96} or median \cite{Krzyzak_et_al01} processing of images, where one seeks a space-filling curve which facilitates the one-dimensional signal processing of multidimensional data, digital halftoning \cite{Velho_Gomes91}, where a space filling curve is sought in order to minimize the effect of false contours, and pattern recognition \cite{Skubalska-Rafajlowicz01}, where it is shown that under certain conditions, the Bayes risk as well as the optimal decision rule are unchanged if instead of the original multidimensional classification problem one transforms the data using a measure-preserving space-filling curve and solves a simpler one-dimensional problem. More applications can be found in multidimensional data query \cite{Asano_et_al97}, \cite{Chung_et_al00} and indexing \cite{Moon_et_al01}, where multidimensional data is stored on a one-dimensional storage device, hence a locality-preserving space-filling curve is sought in order to minimize the number of continuous read operations required to access a multidimensional object, and rendering of three-dimensional graphics \cite{Bogomjakov_Gotsman02}, \cite{Niedermeier_et_al02}, where a rendering sequence which minimizes the number of cache misses is required.

The above applications have already been considered in the literature, and the benefits of not-trivial scanning orders have been proved (see \cite{Tang_et_al04}, or \cite{Memon_et_al95} and \cite{Dafner_et_al00} which we discuss later). Yet, the scandiction problem may have applications that go beyond image scanning alone. For example, consider a robot processing various types of products in a warehouse. The robot identifies a product using a bar-code or an RFID, and processes it accordingly. If the robot could predict the next product to be processed, and prepare for that operation while commuting to the product (e.g., prepare an appropriate writing-head and a message to be written), then the total processing time could be smaller compared to preparing for the operation only after identifying the product. Since different sites in the warehouse my be correlated in terms of the various products stored in them, it is natural to ask what is the optimal path to scan the entire warehouse in order to achieve minimum prediction error and thus minimal processing time. 
 
In \cite{Lempel_Ziv86}, a specific scanning method was suggested by Lempel
and Ziv for the lossless compression of multidimensional data. It was
shown that the application of the incremental parsing algorithm of
\cite{Ziv_Lemp78} on the one dimensional sequence resulting from the
\emph{Peano-Hilbert} scan yields a universal compression algorithm with
respect to all finite-state \emph{scanning and encoding machines}. These results
where later extended in \cite{Weiss_Mannor00} to the probabilistic setting,
where it was shown that this algorithm is also universal for any
stationary Markov random field \cite{Ye_Berg98}. Using the universal
quantization algorithm of \cite{Ziv_85}, the existence of a universal
rate-distortion encoder was also established. Additional results regarding
lossy compression of random fields (via pattern matching) were given in
\cite{Dembo_Kontoyiannis02} and \cite{Kontoyiannis03}. For example, in
\cite{Kontoyiannis03}, Kontoyiannis considered a lossy encoder which encodes
the random field by searching for a $D$-closest match in a given database,
and then describing the position in the database.

While the algorithm suggested in \cite{Lempel_Ziv86} is asymptotically optimal, it may not be the optimal compression algorithm for real life images of sizes such as $256 \times 256$ or $512 \times 512$. In \cite{Memon_et_al95}, Memon \emph{et}.\ \emph{al}.\ considered image compression with a codebook of block scans. Therein, the authors sought a scan which minimizes the zero order entropy of the \emph{difference image}, namely, that of the sequence of differences between each pixel and its preceding pixel along the scan. Since this problem is computationally expensive, the authors aimed for a suboptimal scan which minimizes the sum of absolute differences. This scan can be seen as a minimum spanning tree of a graph whose vertices are the pixels in the image and whose edges weights represent the differences (in gray levels) between each pixel and its adjacent neighbors. Although the optimal spanning tree can be computed in linear time, encoding it may yield a total bit rate which is higher than that achieved with an ordinary raster scan. Thus, the authors suggested to use a \emph{codebook of scans}, and encode each block in the image using the best scan in the codebook, in the sense of minimizing the total loss. 

Lossless compression of images was also discussed by Dafner \emph{et}.\ \emph{al}.\ in \cite{Dafner_et_al00}. In this work, a context-based scan which minimizes the number of edge crossing in the image was presented. Similarly to \cite{Memon_et_al95}, a graph was defined and the optimal scan was represented through a minimal spanning tree. Due to the bit rate required to encode the scan itself the results fall short behind \cite{Lempel_Ziv86} for two-dimensional data, yet they are favorable when compared to applying the algorithm in \cite{Lempel_Ziv86} \emph{to each frame} in a three-dimensional data (assuming the context-based scans for each frame in the algorithm of \cite{Dafner_et_al00} are similar).  

Note that although the criterion chosen by Memon \emph{et}.\ \emph{al}.\ in \cite{Memon_et_al95}, or by Dafner \emph{et}.\ \emph{al}.\ in \cite{Dafner_et_al00}, which is to minimize the sum of cumulative (first order) prediction errors or edge crossings, is similar to the criterion defined in this work, there are two important differences. First, the weights of the edges of the graph should be computed before the computation of the optimal (or suboptimal) scan begins, namely, the algorithm is not sequential in the sense of scanning and prediction in one pass. Second, the weights of the edges can only represent prediction errors of first order predictors (i.e., context of length one), since the prediction error for longer context depends on the scan itself - which has not been computed yet.   
In the context of lossless image coding it is also important to mention the work of Memon \emph{et}.\ \emph{al}.\ in \cite{Memon_et_al00}, where common scanning techniques (such as raster scan, Peano-Hilbert and random scan) were compared in terms of minimal cumulative conditional entropy given a \emph{finite} context (note that for unlimited context the cumulative conditional entropy does not depend on the scanning order, as will be elaborated on later). The image model was assumed to be an isotropic Gaussian random filed. Surprisingly, the results of \cite{Memon_et_al00} show that context-based compression techniques based on limited context may not gain by using Hilbert scan over raster scan. Note that under a different criterion, cumulative \emph{squared} prediction error, the raster scan is indeed optimal for Gaussian fields, as it was shown later in \cite{Mer_Weiss03}, which we discuss next. 

The results of \cite{Lempel_Ziv86} and \cite{Weiss_Mannor00} considered a
specific, data independent scan of the data set. Furthermore, even in the works of Memon \emph{et}.\ \emph{al}.\ \cite{Memon_et_al95} or Dafner \emph{et}.\ \emph{al}.\ \cite{Dafner_et_al00}, where data dependent scanning was considered, only limited prediction methods (mainly, first order predictors) were discussed, and the criterion used was minimal total bit rate of the encoded image. However, for a general predictor, loss function and random field (or individual image), it is not clear what is the optimal scan. This more general scenario was
discussed in \cite{Mer_Weiss03}, where Merhav and Weissman formally
defined the notion of a \emph{scandictor}, a scheme for both scanning and
prediction, as well as that of \emph{scandictability}, the best expected
performance on a data array. The main result in
\cite{Mer_Weiss03} is the fact that if a stochastic field can be
represented autoregressively (under a specific scan $\Psi$)
with a maximum-entropy innovation process, then it is optimally scandicted
in the way it was created (i.e., by the specific scan $\Psi$ and its
corresponding optimal predictor).

While defining the yardstick for analyzing the scandiction problem, the work in \cite{Mer_Weiss03} leaves many open
challenges. As the topic of prediction is rich
and includes elegant solutions to various problems, seeking
analogous results in the scandiction scenario offers plentiful research
objectives.

In Section \ref{sec. Universal Scandiction}, we consider the case where one strives to compete with a finite set $\calF$ of scandictors. Specifically, assume that there exists a probability measure $Q$ which governs the data array. Of course, given the probability measure $Q$ and the scandictor set, one can compute the optimal scandictor in the set (in some sense which will be defined later). However, we are interested in a universal scandictor, which scans the data independently of $Q$, and yet achieves essentially the same performance as the best scandictor in $\calF$ (see \cite{Mer_Fed98} for a complete survey of universal prediction). The reasoning behind the actual choice of the scandictor set $\calF$ is similar to that common in the filtering and prediction literature (e.g., \cite{Gyorfi_Lugosi_Morvai99} and \cite{Weiss_Mer04}). On the one hand, it should be large enough to cover a wide variety of random fields, in the sense that for each random field in the set, at least one scandictor is sufficiently close to the optimal scandictor corresponding to that random field. On the other hand, it should be small enough to compete with, at an acceptable cost of redundancy. 

At first sight, in order to compete successfully with a finite set of scandictors, i.e., construct a universal scandictor, one may try to use known algorithms for learning with expert advice, e.g., the \emph{exponential weighting} algorithm suggested in \cite{Vovk90} or the work which followed it. In this algorithm, each expert is assigned a weight according to its past performance. By decreasing the weight of poorly performing experts, hence preferring the ones proved to perform well thus far, one is able to compete with the best expert, having neither any \emph{a priori} knowledge on the input sequence nor which expert will perform the best. However, in the scandiction problem, as each of the experts may use a different scanning strategy, at a given point in time each scanner might be at a different site, with different sites as its past. Thus, it is not at all guaranteed that one can alternate from one expert to the other. The problem is even more involved when the data is an individual image, as no statistical properties of the data can be used to facilitate the design or analysis of an algorithm. In fact, the first result in Section \ref{sec. Universal Scandiction} is a negative one, stating that indeed in the individual image scenario (or under expected minimum loss in the stochastic scenario), it is not possible to successfully compete with any two scandictors on any individual image. This negative result shows that the scandiction problem is fundamentally different and more challenging than the previously studied problems, such as prediction and compression, where competition with an arbitrary finite set of schemes in the individual sequence setting is well known to be an attainable goal. However, in Theorem \ref{theo. universal finite set scandictability} of Section \ref{sec. Universal Scandiction}, we show that for the class of spatially stationary random fields, and subject to mild conditions on the prediction loss function, one can compete with any finite set of scandictors (under minimum expected loss). Furthermore, in Theorem \ref{theo. existence of an alg achieving U}, our main result in this section, we introduce a universal scandictor for any spatially stationary random field. Section \ref{sec. Universal Scandiction} also includes almost surely analogues of the above theorems for mixing random fields and basic results on cases where universal scandiction of individual images is possible.  

In Section \ref{sec. Bounds}, we derive upper bounds on the excess loss incurred when non-optimal scanners are used, with optimal prediction schemes. Namely, we consider the scenario where one cannot use a universal scandictor (or the optimal scan for a given random field), and instead uses an arbitrary scanning order, accompanied by the optimal predictor for that scan. In a sense, the results of Section \ref{sec. Bounds} can be used to assess the sensitivity of the scandiction performance to the scanning order. Furthermore, in Section \ref{sec. Bounds} we also discuss the scenario where the Peano-Hilbert scanning order is used, accompanied by an optimal predictor, and derive a bound on the excess loss compared to optimal finite state scandiction, which is valid for any individual image and any bounded loss function. Section \ref{sec.conc} includes a few concluding remarks and open problems.

In \cite{Cohen_Weissman_Merhav_II07}, the second part of this two-part paper, we consider sequential decision making for \emph{noisy} data arrays. Namely, the decision maker observes a noisy version of the
data, yet, it is judged with respect to the clean data. As the clean data is not available, two distinct cases are interesting to consider. The first, \emph{scanning and filtering}, is when $Y_{\Psi_t}$ is available in the estimation of $X_{\Psi_t}$, i.e., $F_t$ depends on $Y_{\Psi_1}$ to $Y_{\Psi_{t}}$, where $\{Y\}$ is the noisy data. The second, noisy scandiction, is when the noisy observation at the current site is not available to the decision maker. In both cases, the decision maker cannot evaluate its performance precisely, as $l(x_{\Psi_t},F_t)$ cannot be computed. Yet, many of the results for noisy scandiction are extendable from the noiseless case, similarly as results for noisy prediction were extended from results for noiseless prediction \cite{Weiss_Mer01}. The scanning and filtering problem, however, poses new challenges and requires the use of new tools and techniques. Thus, in \cite{Cohen_Weissman_Merhav_II07}, we formally define the best achievable performance in these cases, derive bounds on the excess loss when non-optimal scanners are used and present universal algorithms. A special emphasis is given on the cases of binary random fields corrupted by a binary memoryless channel and real-valued fields corrupted by Gaussian noise. 

%% file: formulation.tex
\section{Problem Formulation}\label{sec. prob. form.}The following notation will be used throughout this paper.\footnote{For easy reference, we try to follow the notation of
  \cite{Mer_Weiss03} whenever possible.} Let $A$ denote the alphabet, which is either discrete or the real line. Let
$\Omega = A^{\Z^d}$ denote the
space of all possible data arrays in $\Z^d$. Although the results in
this paper are applicable to any $d \geq 1$, for simplicity, we assume from now on that $d=2$. The extension to $d \geq 2$ is straightforward. A probability measure $Q$ on $\Omega$ is stationary
if it is invariant under translations $\tau_i$, where for each $x \in \Omega$ and $i,j \in \Z^2$, $\tau_i(x)_j=x_{j+i}$ (namely, stationarity means shift invariance). Denote by $\calM(\Omega)$ and $\calM_S(\Omega)$ the
spaces of all probability measures and stationary probability measures on $\Omega$,
respectively. Elements of $\calM(\Omega)$, \emph{random fields}, will
be denoted by upper case letters while elements of $\Omega$, \emph{individual
data arrays}, will be denoted by the corresponding lower case.

\begin{figure}
\centering
\includegraphics[scale=0.45]{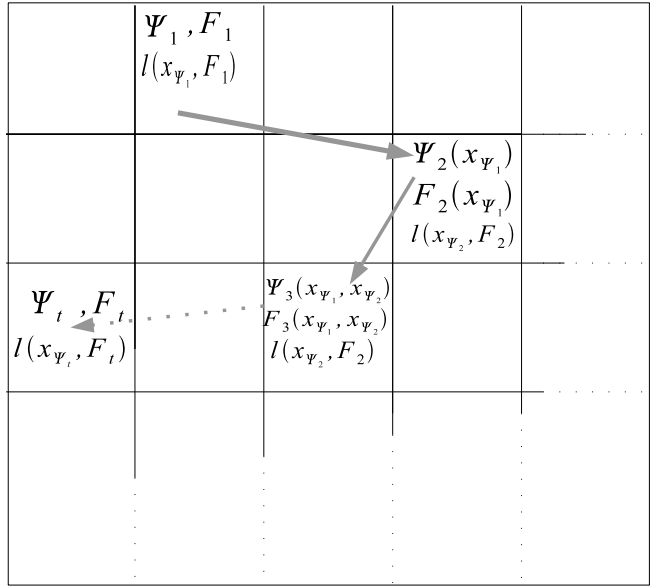}
\caption{A graphical representation of the scandiction process. A scandictor $(\Psi,F)$ first chooses an initial site $\Psi_1$. It then gives its prediction for the value at that site, $F_1$. After observing the true value at $\Psi_1$, it suffers a loss $l(x_{\Psi_1},F_1)$, chooses the next site to be visited, $\Psi_2(x_{\Psi_1})$, gives its prediction for the value at that site, $F_2(x_{\Psi_1})$, and so on.}
\label{fig. scandictor}
\end{figure}

Let $\calV$ denote the set of all finite subsets of $\Z^2$. For $V \in
\calV$, denote by $X_V$ the restrictions of the data array $X$ to $V$. For
$i \in \Z^2$, $X_i$ is the random variable corresponding to $X$ at site
$i$. Let $\calRs$ be the set of all rectangles of the form $V=\Z^2 \cap
([m_1,m_2]\times[n_1,n_2])$. As a special case, denote by $V_n$ the square
$\{0,\ldots,n-1\} \times \{0,\ldots,n-1\}$. For $V \subset \Z^2$, let the interior diameter of
$V$ be
\begin{equation}
R(V) \defined \sup\{r: \exists c \text{ s.t. } B(c,r) \subseteq V\},
\end{equation}
where $B(c,r)$ is a closed ball (under the $l_1$-norm) of radius $r$
centered at $c$. Throughout, $\log(\cdot)$ will denote the natural
logarithm, and entropies will be measured in nats.
\begin{definition}[\cite{Mer_Weiss03}]\label{def. scandictors}
A \emph{scandictor} for a finite set of sites $B \in \calV$ is the following
pair $(\Psi,F)$:
\begin{itemize}
\item $\{\Psi_t\}_{t=1}^{|B|}$ is a sequence of
measurable mappings, $\Psi_t: A^{t-1} \mapsto B$ determining the site
to be visited at time $t$, with the property that
\begin{equation}
\left\{\Psi_1,\Psi_2(x_{\Psi_1}),\Psi_3(x_{\Psi_1},x_{\Psi_2})\ldots,\Psi_{|B|}\left(x_{\Psi_1},\ldots,x_{\Psi_{|B|-1}}\right)\right\}=B,
\quad \forall x \in A^B.
\end{equation}
\item $\{F_t\}_{t=1}^{|B|}$ is a sequence of
measurable predictors, where for each $t$, $F_t: A^{t-1} \mapsto D$ determines the prediction for the site visited at time $t$ based on the observations at previously visited sites, and $D$ is the prediction alphabet.
\end{itemize}
\end{definition}
We allow \emph{randomized scandictors}, namely, scandictors such that
$\{\Psi_t\}_{t=1}^{|B|}$ or $\{F_t\}_{t=1}^{|B|}$ can be chosen randomly
from some set of possible functions. At this point, it is important to
note that scandictors for \emph{infinite} data arrays are not
considered in this paper. Definition \ref{def. scandictors}, and
the results to follow, consider only scandictors for finite
sets of sites, ones which can be viewed merely as a reordering of the
sites in a finite set $B$. We will consider, though, the limit as the
size of the array tends to infinity. A scandictor, such that there exists a finite
set of sites $B$, for which there is no deterministic finite point in
time by which all sites in $B$ are scanned, is not included in the
scope of Definition \ref{def. scandictors}. Figure \ref{fig. scandictor} includes a graphical representation of the scandiction process.

Denote by $L_{(\Psi,F)}(x_{V_n})$ the cumulative loss of $(\Psi,F)$ over $x_{V_n}$,
that is
\begin{equation}\label{eq. cumulative loss}
L_{(\Psi,F)}(x_{V_n}) = \sum_{t=1}^{|V_n|} {l
\left(x_{\Psi_t},F_t(x_{\Psi_1},\ldots,x_{\Psi_{t-1}})\right)},
\end{equation}
where $l:A\times D \rightarrow [0,\infty)$ is a given loss function.
Throughout this paper, we assume that $l(\cdot,\cdot)$ is non-negative and bounded by
$l_{max} < \infty$. The scandictability of a source $Q \in \calM(\Omega)$ on $B \in
\calV$ is defined by
\begin{equation}\label{def. U(l,B)}
U(l,Q_B)=\inf_{(\Psi,F)\in
  \calS(B)}E_{Q_B}\frac{1}{|B|}L_{(\Psi,F)}(X_B),
\end{equation}
where $Q_B$ is the marginal probability measure of $X$ restricted to $B$ and $\calS(B)$ is the set of \emph{all} possible scandictors for
$B$. The scandictability of $Q \in \calM(\Omega)$ is defined by
\begin{equation}\label{def. mer_weiss scandict.}
U(l,Q)=\lim_{n \rightarrow \infty}U(l,Q_{V_n}).
\end{equation}
By \cite[Theorem 1]{Mer_Weiss03}, the limit in \eqref{def. mer_weiss
  scandict.} exists for any $Q \in \calM_S(\Omega)$ and, in fact, for any
sequence $\{B_n\}$ of elements of $\calRs$ for which $R(B_n)\rightarrow \infty$ we have
\begin{equation}\label{eq. existence of the U(l,Q) limit}
U(l,Q)=\lim_{n \rightarrow \infty}U(l,Q_{B_n})=\inf_{B \in \calRs}U(l,Q_B).
\end{equation}
\subsection{Finite-Set Scandictability}
It will be constructive to refer to the \emph{finite
set} scandictability as well. Let $\calF=\{\calF_n\}$ be a sequence of
finite sets of scandictors, where for each $n$, $|\calF_n| = \lambda < \infty$, and
the scandictors in $\calF_n$ are defined
for the finite set of sites $V_n$. A possible scenario is one in
which one has a set of ``scandiction rules'', each of which defines a
unique scanner for each $n$, but all these scanners comply with the same
rule. In this case, $\calF=\{\calF_n\}$ can also be viewed as a finite set $\calF$ which includes sequences of scandictors. For example, $|\calF_n| =2$ for all $n$, where for each $n$, $\calF_n$ includes one scandictor which scans the data row-wise and one which scans the data column-wise. We may also consider cases
in which $|\calF_n|$ increases with $n$ (but remains finite for every finite
$n$). For $Q \in \calM_S(\Omega)$ and $\calF=\{\calF_n\}$, we thus define
the finite set scandictability of $Q$
as the limit
\begin{equation} \label{def. finite set scandictability}
U_{\calF}(l,Q) \defined \lim_{n \rightarrow \infty}\min_{(\Psi,F) \in \calF_n}
E_{Q_{V_n}}\frac{1}{|V_n|}L_{(\Psi,F)}(X_{V_n}),
\end{equation}
if it exists. Observe that the sub-additivity property of the scandictability
as defined in \cite{Mer_Weiss03}, which was fundamental for the existence
of the limit in \eqref{def. mer_weiss scandict.}, does not carry over
verbatim to finite set scandictability. This is for the following reason. Suppose $(\Psi,F) \in \calS$ is the optimal
scandictor for $X_V$ and $(\Psi',F') \in \calS$ is optimal for $X_U$
(assume $V \cap U = \emptyset$). When scanning $X_{V \cup U}$, one may not
be able to apply $(\Psi,F)$ for $X_V$ and then $(\Psi',F')$ for $X_U$, as
this scandictor might not be in $\calS$. Hence, we seek a universal scheme which
competes successfully (in a sense soon to be defined) with a sequence
of finite sets of
scandictors $\{\calF_n\}$, even when the limit in \eqref{def. finite set
scandictability} does not exist.

%% file: universal.tex
\section{Universal Scandiction}\label{sec. Universal Scandiction}
The problem of universal prediction is well studied, with various solutions to both the stochastic setting as well as the individual. In this section, we study the problem of universal \emph{scandiction}. Notwithstanding strongly related to its prediction analogue, we first show that this problem is fundamentally different in several aspects, mainly due to the enormous degree of freedom in choosing the scanning order. Particularly, we first give a negative result, stating that while in the prediction problem it is possible to compete with any finite number of predictors and on every individual sequence, in the scandiction problem one cannot even compete with any two scandictors on a given individual data array. Nevertheless, we show that in the setting of stationary random fields, and under the minimum expected loss criterion, it is possible to compete with any finite set of
scandictors. We then show that the set of \emph{finite-state
  scandictors} is capable of achieving the scandictability of any
spatially stationary source. In Theorem \ref{theo. existence of an alg achieving U}, our main result in this section, we give a universal algorithm which achieves the scandictability of any spatially stationary source.
\subsection{A Negative Result on Scandiction}\label{sec. negative result}
Assume both the alphabet $A$ and the prediction space $D$ are $[0,1]$. Let $l$ be any \emph{non-degenerated} loss function, in the sense that prediction of a bernoulli sequence under it results in a positive expected loss. As an example, squared or absolute error can be kept in mind, though the result below applies to many other loss functions. The following theorem asserts that in the individual image scenario, it is not possible to compete successfully with any two arbitrary scandictors (it is possible, though, to compete with \emph{some} scandictor sets, as proved in Section \ref{subsec. universality for ind. images}).   
\begin{theorem}\label{theo. negative individual result}
Let $A=D=[0,1]$ and assume $l$ is a non-degenerated loss function. There exist two scandictors $(\Psi,F)_1$ and $(\Psi,F)_2$ for $V_n$, such that for any scandictor $(\Psi,F)$ for $V_n$ there exists $x_{V_n}$ for which
\begin{equation}\label{eq. theta in negative result}
L_{(\Psi,F)}(x_{V_n})-\min\{L_{(\Psi,F)_1}(x_{V_n}),L_{(\Psi,F)_2}(x_{V_n})\} =\Theta(|V_n|).
\end{equation}
\end{theorem}

In words, there exist two scandictors such that for any third scandictor, there exists an individual image for which the redundancy when competing with the two scandictors does not vanish. Theorem \ref{theo. negative individual result} marks a fundamental difference between the case where reordering of the data is allowed, e.g., scanning of multidimensional data or even reordering of one-dimensional data, and the case where there is one natural order for the data. For example, using the exponential weighting algorithm discussed earlier, it is easy to show that in the prediction problem (i.e., with no scanning), it is possible to compete with \emph{any finite set} of predictors under the above alphabets and loss functions. Thus, although the
scandiction problem is strongly related to its prediction analogue, the
numerous scanning possibilities result in a substantially richer and
more challenging problem.

Theorem \ref{theo. negative individual result} is a direct application of the lemma below.
%
\begin{lemma}\label{lem. negative stochastic result}
Let $A=D=[0,1]$ and assume $l$ is a non-degenerated loss function. There exist a random field $X_{V_n}$ and two scandictors $(\Psi,F)_1$ and $(\Psi,F)_2$ for $V_n$, such that for any scandictor $(\Psi,F)$ for $V_n$,
\begin{equation}
EL_{(\Psi,F)}(X_{V_n})-E\min\{L_{(\Psi,F)_1}(X_{V_n}),L_{(\Psi,F)_2}(X_{V_n})\} =\Theta(|V_n|).
\end{equation}
\end{lemma}

Lemma \ref{lem. negative stochastic result} gives another perspective on the difference between the scandiction and prediction scenarios. The lemma asserts that when ordering of the data is allowed, one cannot achieve a vanishing redundancy with respect to the \emph{expected value of the minimum} among a set of scandictors. This should be compared to the prediction scenario (no reordering), where one can compete successfully not only with respect to the minimum of the expected losses of all the predictors, but also with respect to the expected value of the minimum (for example, see \cite[Corollary 1]{Weiss_et_al07}). The main result of this section, however, is that for any stationary random field and under mild conditions on the loss function, one can compete successfully with any finite set of scandictors when the performance criterion is the \emph{minimum expected loss}. 
\begin{proof}{(Lemma \ref{lem. negative stochastic result})}
Let $Y_{V_n}$ be a random field such that $Y(1,1)$ is distributed uniformly on $[0,1]$, and $Y_{V_n} \setminus Y(1,1)=Y(1,2), \ldots, Y(1,n),Y(2,1),Y(2,2),\ldots,Y(n,n)$ are simply the first $n^2-1$ bits in the binary representation of $Y(1,1)$ (ordered row-wise). Note that $Y_{V_n} \setminus Y(1,1)$ are i.i.d.\ unbiased bits, yet conditioned on $Y(1,1)$, they are deterministic and known. Assume now that $X_{V_n}$ is a random cyclic shift of $Y_{V_n}$, in the same row-wise order $Y_{V_n}$ was created. 

For concreteness, we assume the squared error loss function. In this case, it is easy to identify the constant of the $\Theta(\cdot)$ expression in \eqref{eq. theta in negative result}. However, the computations below are easily generalized to other non-degenerated loss functions. We first show that the expected cumulative squared error of any scandictor on $X_{V_n}$ is at least $(n^2+1)/8$, as the expected number of steps until the real valued site is located is $(n^2+1)/2$, with a loss of $1/4$ until that time. More specifically, let $J$ be the random number of cyclic shifts, that is, $J$ is uniformly distributed on $\{0,1,\ldots,n^2-1\}$. For fixed $j$, let $G$ be the random index such that $\Psi_G$ is the real-valued $X$ (i.e., $G$ is the time the real valued random variable is located by the scanner $\Psi$). Let $\phi_s$ denote the
Bayes envelope associated with the squared error loss, i.e.,
\begin{equation}\label{def. phi_S}
\phi_s(p)=\min_{q\in[0,1]}[(1-p)q^2+p(q-1)^2].
\end{equation}
For any scandictor $(\Psi,F)$, we have, 
\begin{eqnarray}
EL_{(\Psi,F)}(X_{V_n}) &=& E_J E\left\{\sum_{i=1}^{n^2}{\left(X_{\Psi_i}-F_i(X_{\Psi_1}^{\Psi_{i-1}})\right)^2} \Bigg|j \right\}
\nonumber\\
&=&E_J E\left\{\sum_{i=1}^{G}{\left(X_{\Psi_i}-F_i(X_{\Psi_1}^{\Psi_{i-1}})\right)^2}+\sum_{i=G+1}^{n^2}{\left(X_{\Psi_i}-F_i(X_{\Psi_1}^{\Psi_{i-1}})\right)^2} \Bigg|j \right\}
\nonumber\\
& \ge &E_J E\left\{\sum_{i=1}^{G}{\left(X_{\Psi_i}-F_i(X_{\Psi_1}^{\Psi_{i-1}})\right)^2}\Bigg|j \right\}
\nonumber\\
& \ge & E_J E\left\{\sum_{i=1}^{G}{\phi_s \left(P(X_{\Psi_i}|X_{\Psi_1}^{\Psi_{i-1}})\right)}\Bigg|j \right\}
\nonumber\\
& = & E_J E\left\{G{\phi_s \left(P(X_{\Psi_1}=1)\right)}\Bigg|j \right\}
\nonumber\\
& = & \frac{1}{4}E\{G\}
\nonumber\\
&=& \frac{n^2+1}{8}.\label{eq. proof of negative result}
\end{eqnarray} 
On the other hand, consider the expected \emph{minimum of the losses} of the following two scandictors: $(\Psi,F)_1$ which scandicts $X_{V_n}$ row-wise from $X(1,1)$ to $X(n,n)$, and $(\Psi,F)_2$ which scandicts $X_{V_n}$ row-wise from $X(n,n)$ to $X(1,1)$. Using the same method as in \eqref{eq. proof of negative result}, it is possible to show that this expected loss is \emph{smaller} than $n^2/16 + o(n^2)$, as the expected number of steps until \emph{the first} locates the real-valued site is $(n^2+1)^2/(4n^2)$, after which zero loss is incurred. This is since once the real-valued site is located, the rest of the values can be calculated by the predictor by cyclic shifting the binary representation of the real-valued pixel.  This completes the proof.
\end{proof}
\begin{proof}{(Theorem \ref{theo. negative individual result})} 
By Lemma \ref{lem. negative stochastic result}, there exists a stochastic setting under which the expected minimum of the losses of two scandictors is smaller than the expected loss of any single scandictor. Thus, for any scandictor there exists an individual image on which it cannot compete successfully with the two scandictors.
\end{proof}
\subsection{Universal Scandiction With Respect to Arbitrary Finite Sets}
As mentioned in Section \ref{sec. intro.}, straightforward implementation of the exponential weighting algorithm is not feasible, since one may not be able to alternate from one expert to the other at wish. However, the exponential weighting algorithm was found useful in several lossy
source coding works such as Linder and Lugosi \cite{Linder_Lugosi01}, Weissman
and Merhav \cite{Weiss_Mer02}, Gyorgy \emph{et}.\ \emph{al}.\
\cite{Gyorgy_Linder_Lugosi04} and the derivation of sequential strategies
for loss functions with memory \cite{Mer_Orden_Serou_Weinb02}, all of which confronted a similar problem. A common
method used in these works, is the alternation of experts only once every block
of input symbols, necessary to bear the price of this change (e.g.,
transmitting the description of the chosen quantizer
\cite{Linder_Lugosi01}-\cite{Gyorgy_Linder_Lugosi04}). Thus, although the difficulties in these examples differ from those we confront here, the solution suggested therein, which is to persist on using the same expert for a significantly long block of data before alternating it, was found useful in our universal scanning problem. 

Particularly, we divide the data array into smaller blocks and alternate
scandictors only each time a new block of data is to be scanned. Unlike the
case of sequential prediction dealt with in \cite{Mer_Orden_Serou_Weinb02},
here the scandictors must be \emph{restarted} each time a new block is scanned,
as it is not at all guaranteed that all the scandictors scan the data in the
same (or any) block-wise order (i.e., it is not guaranteed that a scandictor
for $V_n$ divides the array to sub-blocks of size $m \times m$ and scans each
of them separately). Hence, in order to prove that it is possible to compete
with the best scandictor at each stage $n$, we go through two phases. In the
first, we prove that an exponential weighting algorithm may be used to compete
with the best scandictor among those operating in a block-wise order. This part
of the proof will refer to any given data array (deterministic scenario). In
the second phase, we use the stationarity of the random field to prove that a
block-wise scandictor may perform essentially as well as one scanning the data
array as a whole. The following theorem stands at the basis of our results,
establishing the existence of a universal scandictor which competes
successfully with any finite set of scandictors.
%
\begin{theorem} \label{theo. universal finite set scandictability}
Let $X$ be a stationary random field with a probability measure $Q$. Let
$\calF = \{\calF_n\}$ be an arbitrary sequence of scandictor sets, where $\calF_n$ is a set of scandictors for $V_n$ and $|\calF_n| = \lambda < \infty$ for all $n$. Then, there
exists a sequence of scandictors
$\{(\hat{\Psi},\hat{F})_n\}$, where $(\hat{\Psi},\hat{F})_n$ is a scandictor for $V_n$, independent of $Q$, for which
\begin{equation} \label{eq. inq. between the liminf}
\liminf_{n \rightarrow \infty}
E_{Q_{V_n}}E\frac{1}{|V_n|}L_{(\hat{\Psi},\hat{F})_n}(X_{V_n}) \leq
\liminf_{n \rightarrow \infty}\min_{(\Psi,F) \in \calF_n}
E_{Q_{V_n}}\frac{1}{|V_n|}L_{(\Psi,F)}(X_{V_n})
\end{equation}
for any $Q \in \calM_S(\Omega)$, where the inner expectation in the
l.h.s. of \eqref{eq. inq. between the
liminf} is due to the possible randomization in
$(\hat{\Psi},\hat{F})_n$.
\end{theorem}
Before we prove Theorem \ref{theo. universal finite set scandictability}, let
us discuss an ``individual image'' type of result, which will later be the
basis of the proof. Let $x_{V_n}$ denote an individual $n \times n$ data array.
For $m<n$, define $K \defined \Ceil[\frac{n}{m}]-1$. We divide $x_{V_n}$ into
$K^2$ blocks of size $m\times m$ and $2K+1$ blocks of possibly smaller size.
Denote by $x^i$, $0 \leq i \leq (K+1)^2-1$ the $i$'th block under some fixed
scanning order of the blocks. Since we will later see that this scanning order
is irrelevant in this case, assume from now on that it is a (continuous) raster
scan from the upper left corner. That is, the first line of blocks is scanned
left to right, the second line is scanned right to left, and so on. We will
refer to this scan simply as ``raster scan".

As mentioned, the suggested algorithm scans the data in $x_{V_n}$ block-wise,
that is, it does not apply any of the scandictors in $\calF_n$, only
scandictors from $\calF_m$. Omitting $m$ for convenience, denote by $L_{j,i}$
the cumulative loss of $(\Psi,F)_j \in \calF_m$ after scanning $i$ blocks,
where $(\Psi,F)_j$ is \emph{restarted} after each block, namely, it scans each
block separately and independently of the other blocks. Note that $L_{j,i} =
\sum_{l=0}^{i-1}{L_j(x^l)}$ and that for $i=0$, $L_{j,i}=0$ for all $j$. Since
we assumed the scandictors are capable of scanning only square blocks, for the
$2K+1$ possibly smaller (and not square) blocks the loss may be $l_{max}$
throughout. For $\eta>0$, and any $i$ and $j$, define
\begin{equation}
P_i\left(j|\{L_{j,i}\}_{j=1}^{\lambda}\right)=\frac{e^{-\eta
L_{j,i}}}{\sum_{j=1}^{\lambda}{e^{-\eta L_{j,i}}}},
\end{equation}
where $\lambda=|\calF_m|$. We offer the following algorithm for a block-wise
scan of the data array $x$. For each $0 \leq i \leq (K+1)^2-1$, after scanning
$i$ blocks of data, the algorithm computes
$P_i\left(j|\{L_{j,i}\}_{j=1}^{\lambda}\right)$ for each $j$. It then randomly
selects a scandictor according to this distribution, independently of its
previous selections, and uses this scandictor as its output for the $(i+1)$-st
block. Namely, the universal scandictor $(\hat{\Psi},\hat{F})_n$, promised by
Theorem \ref{theo. universal finite set scandictability}, is the one which
divides the data to blocks, performs a raster scan of the data block-wise, and
uses the above algorithm to decide which scandictor out of $\calF_m$ to use for
each block.

It is clear that both the block size and the number of blocks should tend to
infinity with $n$ in order to achieve meaningful results. Thus, we require the
following: a. $m=m(n)$ tends to infinity, but strictly slower than $n$, i.e.,
$m(n)=o(n)$. b. $m(n)$ is an integer-valued monotonically increasing function,
such that for each $K \in \Z$ there exists $n$ such that $m(n)=K$. The results
are summarized in the following two propositions, the first of which asserts
that for $m(n)=o(n)$, vanishing redundancy is indeed possible, while the second
asserts that under slightly stronger requirements on $m(n)$, this is also true
in the a.s.\ sense (with respect to the random selection of the scandictors in
the algorithm).
%
\begin{proposition}\label{prop. compete with a set of scan. block-wise}
Let $L_{alg}(x_{V_n})$ be the cumulative loss of the proposed algorithm on
$x_{V_n}$, and denote by $\bar{L}_{alg}(x_{V_n})$ its expected value, where the
expectation is with respect to the randomized scandictor selection of the
algorithm. Let $L_{min}$ denote the cumulative loss of the best scandictor in
$\calF_m$, operating block-wise on $x_{V_n}$. Assume $|\calF_m|=\lambda$. Then, for any $x_{V_n}$,
\begin{equation} \label{eq. regret in compete with a set of scan. block-wise}
\bar{L}_{alg}(x_{V_n}) - L_{min}(x_{V_n}) \leq m(n)(n+m(n))\sqrt{\log
\lambda}\frac{l_{max}}{\sqrt{2}}.
\end{equation}
\end{proposition}
%
\begin{proposition}\label{prop. compete with a set of scan. block-wise, a.s.}
Assume $m(n)=o\left(n^{1/3}\right)$. Then, for any image $x_{V_n}$, the difference between the 
normalized cumulative loss of the proposed algorithm and that of the
best scandictor in $\calF_m$, operating block-wise, converges to $0$ with probability $1$ with respect to the randomized scandictor selection of the
algorithm.
\end{proposition}
The proofs of Propositions \ref{prop. compete with a set of
  scan. block-wise} and \ref{prop. compete with a set of
  scan. block-wise, a.s.} are rather technical and are based on the very same methods used
in \cite{Hauss_Kivi_Warm98} and \cite{Mer_Orden_Serou_Weinb02}. See
Appendices \ref{app. proof of prop. comp. with a set} and
\ref{app. proof of prop. comp. with a
  set, a.s.} for the details.

On the more technical side, note that the suggested algorithm has ``finite horizon," that is, one has to know the size of the image in order to
divide it to blocks, and only then can the exponential weighting
algorithm be used. It is possible to extend the algorithm to infinite horizon. The
essence of this generalization is in dividing the infinite image into blocks of
exponentially growing size\footnote{For example, take four
blocks of size $l \times l$, then three of size $2l \times 2l$, and so
on.}, and to apply the finite horizon algorithm for each block. We may now proceed to the proof of Theorem \ref{theo. universal finite set
scandictability}.
\begin{proof}[Proof of Theorem \ref{theo. universal finite set
    scandictability}.] Since the result of Proposition \ref{prop.
compete with a set of scan. block-wise} applies to any individual data
array, it certainly applies after taking the expectation with respect to
$Q$. Therefore,
\begin{equation}\label{eq. diff between alg and min under expect.}
E_{Q_{V_n}}\frac{1}{n^2}\bar{L}_{alg} - E_{Q_{V_n}}\frac{1}{n^2}L_{min} \leq
\frac{m(n)}{n}l_{max}\sqrt{2\log\lambda}.
\end{equation}
However, remember that we are not interested in competing with
$E_{Q_{V_n}}\frac{1}{n^2}L_{min}$, as this is the performance of the best
\emph{block-wise} scandictor. We wish to compete with the best scandictor
operating on the \emph{entire} data array $X_{V_n}$, that is, on the whole
image of size $n\times n$. We have
\begin{eqnarray}
  E_{Q_{V_n}}\frac{1}{n^2}L_{min} &=&
E_{Q_{V_n}}\frac{1}{n^2}\min_{(\Psi,F)_j \in \calF_{m(n)}}
\sum_{i=1}^{(K+1)^2}{L_j(X^i)}
\nonumber\\
& \leq & \min_{(\Psi,F)_j \in \calF_{m(n)}}
E_{Q_{V_n}}\frac{1}{n^2}\sum_{i=1}^{(K+1)^2}{L_j(X^i)}
\nonumber\\
& \stackrel{(a)}{\leq} & \min_{(\Psi,F)_j \in \calF_{m(n)}} \frac{1}{n^2}\cdot
\bigg[K^2 E_{Q_{V_n}}L_j(X^0)
\nonumber\\
&&+2Km(n)(n-Km(n))l_{max} + (n-Km(n))^2l_{max} \bigg]
\nonumber\\
& \leq & \min_{(\Psi,F)_j \in \calF_{m(n)}} E_{Q_{V_n}}\frac{1}{m(n)^2}L_j(X^0)
+ 2\frac{m(n)}{n} l_{max}, \label{eq. diff between min and opt under expect.}
\end{eqnarray}
where $(a)$ follows from the stationarity of $Q$, the assumption that each
$(\Psi,F)_j$ operates in the same manner on each $m(n) \times m(n)$ block, no
matter what its coordinates are, and the fact that each $(\Psi,F)_j$ may incur
maximal loss on non-square rectangles. From \eqref{eq. diff between alg and min
under expect.} and \eqref{eq. diff between min and opt under expect.}, we have
\begin{eqnarray} \label{eq. Lalg vs. Lmin before the limit}
E_{Q_{V_n}}\frac{1}{n^2}\bar{L}_{alg} &\leq&
E_{Q_{V_n}}\frac{1}{m(n)^2}L_{j^*(m(n))}(X^0) + 2\frac{m(n)}{n}l_{max}
 + \frac{m(n)}{n}l_{max}\sqrt{2\log\lambda}
\nonumber\\
&=& E_{Q_{V_n}}\frac{1}{m(n)^2}L_{j^*(m(n))}(X^0) +
O\left(\frac{m(n)}{n}\sqrt{\log\lambda}\right)
\end{eqnarray}
where $(\Psi,F)_{j^*(m(n))}$ is the scandictor achieving the minimum in
\eqref{eq. diff between min and opt under
  expect.}. Finally, by our assumptions on $\{m(n)\}$, we
have
\begin{eqnarray}
\inf_{k \geq n}\left\{E_{Q_{V_k}}\frac{1}{k^2}\bar{L}_{alg}\right\} &\leq&
\inf_{k \geq n}\left\{E_{Q_{V_k}}\frac{1}{m(k)^2}L_{j^*(m(k))}(X^0) +
\frac{m(k)}{k}l_{max}(2+\sqrt{2\log\lambda})\right\}
\nonumber\\
&\leq& \inf_{k \geq
n}\left\{E_{Q_{V_k}}\frac{1}{m(k)^2}L_{j^*(m(k))}(X^0)\right\} +
\frac{m(n)}{n}l_{max}(2+\sqrt{2\log\lambda})
\nonumber\\
&\leq& \inf_{k \geq
n}\left\{E_{Q_{V_k}}\frac{1}{k^2}L_{j^*(k)}(X_{V_k})\right\} +
\frac{m(n)}{n}l_{max}(2+\sqrt{2\log\lambda}).\label{eq. before the limit in
main theorem proof}
\end{eqnarray}
Taking the limit as $n \rightarrow \infty$ and using the fact that $m(k)/k \to
0$ together with the arbitrariness of $k$, gives:
\begin{equation}
\liminf_{n \rightarrow \infty} E_{Q_{V_n}}\frac{1}{n^2}\bar{L}_{alg} \leq
\liminf_{n \rightarrow \infty}
E_{Q_{V_n}}\frac{1}{n^2}L_{j^*(n)}(X_{V_n}),
\end{equation}
which completes the proof of \eqref{eq. inq. between the
  liminf}.
\end{proof}
It is evident from \eqref{eq. regret in compete with a set of scan. block-wise} and \eqref{eq. before the limit in main theorem proof} that although the results of Theorem \ref{theo. universal finite set
    scandictability} and Proposition \ref{prop. compete with a set of scan. block-wise} are formulated for fixed $\lambda < \infty$ (the cardinality of the scandictor set), these results hold for the more general case of $\lambda = \lambda(n)$, as long as the redundancy vanishes, i.e., as long as $m(n)=o(n)$ and $\lambda(n)$ is such that $\frac{m(n)}{n}\sqrt{\log \lambda} \to 0$ when $n \to \infty$. The requirement that $\lambda(n) = o\Big(e^\frac{n^2}{m(n)^2}\Big)$ still allows very large scandictor sets, especially when $m(n)$ grows slowly with $n$. Furthermore, it is evident from equation \eqref{eq. Lalg vs. Lmin before the limit} that whenever the redundancy vanishes, the statement of Theorem \ref{theo. universal finite set scandictability} is valid with $\limsup$ as well ,i.e.,
\begin{equation} \label{eq. inq. between the limsup}
\limsup_{n \rightarrow \infty}
E_{Q_{V_n}}E\frac{1}{|V_n|}L_{(\hat{\Psi},\hat{F})_n}(X_{V_n}) \leq
\limsup_{n \rightarrow \infty}\min_{(\Psi,F) \in \calF_n}
E_{Q_{V_n}}\frac{1}{|V_n|}L_{(\Psi,F)}(X_{V_n}).
\end{equation}
\subsection{Finite-State Scandiction}
Consider now the set of finite-state scandictors, very similar to
the set of finite-state encoders described in \cite{Lempel_Ziv86}. At time
$t=1$, a \emph{finite-state scandictor} starts at an arbitrary initial site $\Psi_1$, with an arbitrary
initial state $s_0 \in S$ and gives $F(s_0)$ as its prediction for
$x_{\Psi_1}$. Only then it observes $x_{\Psi_1}$. After observing
$x_{\Psi_i}$, it computes its next state, $s_i$, according to $s_i =
g(s_{i-1},x_{\Psi_i})$ and advances to the next site,
$x_{\Psi_{i+1}}$, according to $\Psi_{i+1} =
\Psi_i + d(s_i)$, where $g:S \times A \mapsto S$ is the next state
function and $d:S \mapsto B$ is the displacement function, $B
\subset \Z^2$ denoting a fixed finite set of possible relative displacements. It then gives its prediction $F(s_i)$ to the value $x_{\Psi_{i+1}}$. Similarly to \cite{Lempel_Ziv86}, we assume the alphabet $A$
includes an additional ``End of File'' (EoF) symbol to mark the image
edges. The following lemma and the theorem which follows establish the
fact that the set of finite-state scandictors is indeed rich enough
to achieve the scandictability of any stationary source, yet not too
rich to compete with.
\begin{lemma}\label{lem. FSM suffice for stat. sources}
Let $\calF_{S}=\{(\Psi,F)_j\}$ be the set of
all finite-state scandictors with at most $S$ states. Then, for any $Q \in \calM_S(\Omega)$,
\begin{equation}\label{eq. FSM suffice for stat. sources}
\lim_{S \to \infty}U_{\calF_{S}}(l,Q) = U(l,Q).
\end{equation}
That is, the scandictability of any spatially stationary source is
asymptotically achieved with finite-state scandictors.
\end{lemma}
\begin{proof}
Take $B=V_m$ and let $(\bar{\Psi},\bar{F})_m$ be the achiever of the infimum in \eqref{def. U(l,B)}. That is,
\begin{equation}\label{eq. epsilon achiever}
E_{Q_{V_m}}\frac{1}{m^2}L_{(\bar{\Psi},\bar{F})_m}(X_{V_m}) \leq \inf_{(\Psi,F)
\in \calS(V_m)}E_{Q_{V_m}}\frac{1}{m^2}L_{(\Psi,F)}(X_{V_m}).
\end{equation}
Since $V_m$ is a rectangle of size $m \times m$, the scandictor
$(\bar{\Psi},\bar{F})_m$ is certainly implementable with a finite-state machine
having $S(m) < \infty$ states. In other words, since $V_m$ is finite,
\emph{any} scanning rule $\Psi_t: A^{t-1} \mapsto B$ and \emph{any} prediction
rule $F_t: A^{t-1} \mapsto A$ can be implemented with a finite-state machine
having at most $\tilde{S}(m)=A^{m^2} \times m^2$ states, where in a
straightforward implementation $A^{m^2}$ states are required to account for all
possible inputs and $m^2$ states are required to implement a counter.

Now, for an $n \times n$ image (assuming now that $m$ divides $n$, as dealing
with the general case can be done in the exact same way as \eqref{eq. diff
between min and opt under expect.}), we take $(\bar{\Psi'},\bar{F'})_n$ to be the
scandictor which scans the image in the block-by-block raster scan described
earlier, applying $(\bar{\Psi},\bar{F})_m$ to each $m \times m$ block.
Namely, $\bar{\Psi'}$ scans all the blocks in the first $m$ lines left-to-right
until it reaches an EoF symbol, then moves down $m$ lines, scans all the blocks
right-to-left until an EoF is reached, and so on. The predictor $\bar{F'}$
simply implements $\bar{F}$ for each block separately, i.e., it resets to its
initial values at the beginning of each block. It is clear that the scanner
$\bar{\Psi'}$ is implementable with a finite-state machine having
$S(m)=\tilde{S}(m)+2 < \infty$ states and thus $(\bar{\Psi'},\bar{F'}) \in
\calF_{S(m)}$.

From the stationarity of $Q$, we have
\begin{eqnarray}
\inf_{(\Psi,F) \in \calS(V_n)}E_{Q_{V_n}}\frac{1}{n^2}L_{(\Psi,F)}(X_{V_n})
&\leq& \min_{(\Psi,F) \in
\calF_{S(m)}}E_{Q_{V_n}}\frac{1}{n^2}L_{(\Psi,F)}(X_{V_n})
\nonumber\\
&\leq& E_{Q_{V_n}}\frac{1}{n^2}L_{(\bar{\Psi'},\bar{F'})_n}(X_{V_n})
\nonumber\\
&=& E_{Q_{V_m}}\frac{1}{m^2}L_{(\bar{\Psi},\bar{F})_m}(X_{V_m})
\nonumber\\
&\leq& \inf_{(\Psi,F) \in
\calS(V_m)}E_{Q_{V_m}}\frac{1}{m^2}L_{(\Psi,F)}(X_{V_m}).
\end{eqnarray}
Taking the limits $\limsup_{n \to \infty}$ and $\liminf_{n \to \infty}$, by \eqref{eq. existence of the U(l,Q) limit}, we have
\begin{eqnarray}
U(l,Q) &\leq& \limsup_{n \to \infty}\min_{(\Psi,F) \in
  \calF_{S(m)}}E_{Q_{V_n}}\frac{1}{n^2}L_{(\Psi,F)}(X_{V_n})
\nonumber\\
&\leq& \inf_{(\Psi,F) \in
\calS(V_m)}E_{Q_{V_m}}\frac{1}{m^2}L_{(\Psi,F)}(X_{V_m})
\end{eqnarray}
and
\begin{eqnarray}
U(l,Q) &\leq& \liminf_{n \to \infty}\min_{(\Psi,F) \in
  \calF_{S(m)}}E_{Q_{V_n}}\frac{1}{n^2}L_{(\Psi,F)}(X_{V_n})
\nonumber\\
&\leq& \inf_{(\Psi,F) \in
\calS(V_m)}E_{Q_{V_m}}\frac{1}{m^2}L_{(\Psi,F)}(X_{V_m}).
\end{eqnarray}
The proof is completed (including the existence of the limit in the l.h.s. of
\eqref{eq. FSM suffice for stat. sources}) by taking $m$ to infinity, applying
\eqref{eq. existence of the U(l,Q) limit}, and remembering that
$U_{\calF_S}(l,Q)$ is monotone in $S$, thus the convergence of the sub-sequence
$\{U_{\calF_{S(m)}}(l,Q)\}_{m=1}^{\infty}$ implies the convergence of the
sequence $\{U_{\calF_S}(l,Q)\}_{S=1}^{\infty}$).
\end{proof}
In words, Lemma \ref{lem. FSM suffice for stat. sources} asserts that for any
$m$, finite-state machines attain the $m \times m$ Bayesian scandictability for
any stationary random field. Note that the reason such results are
accomplishable with FSMs is their ability to scan the entire data, block by
block, with a machine having no more than $S(m)$ states, regardless of the size
of the complete data array. The number of the states depends only on the block
size.
\subsection{A Universal Scandictor for Any Stationary Random Field}
We now show that a
universal scandictor which competes successfully with all finite-state
machines of the form given in the proof of Lemma \ref{lem. FSM suffice
  for stat. sources}, does exist and can, in fact, be implemented
using the exponential weighting algorithm. In order to show that we assume that the alphabet $A$ is finite and the prediction space $D$ is either finite or bounded (such as the $|D|-1$ simplex of probability measures on $D$). In the latter case we further assume that $l(x,F)$ is Lipschitz in its second argument for all $x$, i.e, there exists a constant $c$ such that for all $x$, $F$ and $\epsilon$ we have $|l(x,F)-l(x,F+\epsilon)| \leq c|\epsilon|$. The following theorem establishes, under the above assumptions, the existence of a universal scandictor for all stationary random fields.
%
\begin{theorem}\label{theo. existence of an alg achieving U}
Let $X$ be a stationary random field over a finite alphabet $A$ and a probability measure $Q$. Let the prediction space $D$ be either finite or bounded (with $l(x,F)$ then being Lipschitz in its second argument). Then, there
exists a sequence of scandictors
$\{(\Psi,F)_n\}$, independent of $Q$, for which
\begin{equation}\label{eq. achieving U}
\lim_{n \rightarrow \infty}
E_{Q_{V_n}}E\frac{1}{|V_n|}L_{(\Psi,F)_n}(X_{V_n}) = U(l,Q)
\end{equation}
for any $Q \in \calM_S(\Omega)$, where the inner expectation in the
l.h.s. of \eqref{eq. achieving U} is due to the possible randomization in
$(\Psi,F)_n$.
\end{theorem}
%
\begin{proof}
Assume first that the range $D$ of the predictors $\{F_t\}$ is finite. Consider
the exponential weighting algorithm described in the proof of Theorem
\ref{theo. universal finite set scandictability}, where at each $m(n) \times
m(n)$ block the algorithm computes the cumulative loss of \emph{every} possible
scandictor for an $m(n) \times m(n)$ block, then chooses the best scandictor
(according to the exponential weighting regime described therein) as its output
for the next block. By \eqref{eq. Lalg vs. Lmin before the limit}, we have
\begin{equation}\label{eq. redundancy in the second proof}
E_{Q_{V_n}}\frac{1}{n^2}\bar{L}_{alg} \leq \min_{(\Psi,F) \in
\calS(V_{m(n)})}E_{Q_{V_{m(n)}}}\frac{1}{m(n)^2}L_{(\Psi,F)}(X^0) +
O\left(\frac{m(n)}{n}\sqrt{\log\lambda}\right),
\end{equation}
where $\calS(V_{m(n)})$ is the set of all possible scandictors on $m(n) \times
m(n)$ and $\lambda$ is the size of that set. Since
$\lambda=\lambda\left(m(n)\right)$, all that is left to check is that the
$O\left(\frac{m(n)}{n}\sqrt{\log\lambda}\right)$ expression indeed decays to
zero as $n$ tends to infinity.

Indeed, the number of possible scanners for a field $B$ over an alphabet $A$ is
\begin{eqnarray}
\left|\left\{ \{\Psi_t\}_{t=1}^{|B|}, \Psi_t:A^{t-1} \mapsto B\right\}\right| &=& \prod_{k=0}^{|B|}(|B|-k)^{|A|^k}
\nonumber\\
&\leq& (|B|!)^{|A|^{|B|}},
\end{eqnarray}
while the number of possible predictors is
\begin{eqnarray}
\left|\left\{ \{F_t\}_{t=1}^{|B|}, F_t:A^{t-1} \mapsto D\right\}\right| &=& \prod_{k=1}^{|B|}|D|^{|A|^k}
\nonumber\\
&\leq& |D|^{|B||A|^{|B|-1}}.
\end{eqnarray}
Thus, using the Stirling approximation, $\log k! \approx k\log k$, in the sense that $\lim_{k \to \infty} \frac{\log k!}{k \log k} = 1$, we have
\begin{eqnarray}
\frac{m(n)}{n}\sqrt{\log\lambda} &\leq& \frac{m(n)}{n} \sqrt{
\log\left[(m(n)^2!)^{|A|^{m(n)^2}} |D|^{m(n)^2|A|^{m(n)^2-1}}\right]}
\nonumber\\
&\approx& \frac{m(n)}{n} \sqrt{2|A|^{m(n)^2}m(n)^2\log m(n) +
m(n)^2|A|^{m(n)^2-1}\log |D|}
\nonumber\\
&\approx& \frac{m(n)^2}{n} \sqrt{|A|^{m(n)^2}\log m(n)},
\end{eqnarray}
which decays to zero as $n \to \infty$ for any $m(n) = o(\sqrt{\log n})$.
Namely, for $m(n) = o(\sqrt{\log n})$, equation \eqref{eq. redundancy in the
second proof} results in
\begin{equation}\label{eq. a.s. proof liminf}
\liminf_{n \to \infty}E_{Q_{V_n}}\frac{1}{n^2}\bar{L}_{alg} \leq \liminf_{n \to \infty}\min_{(\Psi,F) \in
\calS(V_{m(n)})}E_{Q_{V_{m(n)}}}\frac{1}{m(n)^2}L_{(\Psi,F)}(X^0),
\end{equation}
and
\begin{equation}\label{eq. a.s. proof limsup}
\limsup_{n \to \infty}E_{Q_{V_n}}\frac{1}{n^2}\bar{L}_{alg} \leq \limsup_{n \to \infty}\min_{(\Psi,F) \in
\calS(V_{m(n)})}E_{Q_{V_{m(n)}}}\frac{1}{m(n)^2}L_{(\Psi,F)}(X^0).
\end{equation}
Since $m(n) \to \infty$ as $n \to \infty$, by \cite{Mer_Weiss03} the limit $\lim_{n \to \infty}\min_{(\Psi,F) \in
\calS(V_{m(n)})}E_{Q_{V_{m(n)}}}\frac{1}{m(n)^2}L_{(\Psi,F)}(X^0)$ exists and equals the scandictability of the source, $U(l,Q)$. However, by definition, $U(l,Q)$ is the best achievable scandiction performance for the source $Q$, hence,
\begin{equation}\label{eq. a.s. proof liminf vs U}
\liminf_{n \to \infty}E_{Q_{V_n}}\frac{1}{n^2}\bar{L}_{alg} \geq U(l,Q),
\end{equation}
which results in
\begin{equation}\label{eq. bottom line of second proof}
\lim_{n \to \infty}E_{Q_{V_n}}\frac{1}{n^2}\bar{L}_{alg} = U(l,Q).
\end{equation}

For the case of infinite (but bounded) range $D$, similarly to \cite{Weiss_Mer04}, we use the fact that the loss function $l$ is Lipschitz and take an $\epsilon$-approximation of $D$. We thus have
\begin{multline}\label{eq. redundancy in the second proof - with infinite D}
E_{Q_{V_n}}\frac{1}{n^2}\bar{L}_{alg} \leq \min_{(\Psi,F) \in
\calS(V_{m(n)})}E_{Q_{V_{m(n)}}}\frac{1}{m(n)^2}L_{(\Psi,F)}(X^0)
\\
+ c m(n)^2 \epsilon\left(m(n)\right)+
O\left(\frac{m(n)}{n}\sqrt{\log\lambda}\right)
\end{multline}
for some constant $c$. Choosing $\epsilon\left(m(n)\right) = \frac{1}{m(n)^4}$ results in
$|D|=\frac{1}{2}m(n)^4$, hence  $\frac{m(n)}{n}\sqrt{\log\lambda}$ still decays
to zero for any $m(n) = O(\sqrt{\log n})$ and \eqref{eq. bottom line of second
proof} is still valid.
\end{proof}
Note that the proof of Theorem \ref{theo. existence of an alg achieving U} does not use the well established theory of universal prediction. Instead, the exponential weighting algorithm is used for all possible scans (within a block) as well as \emph{all possible predictors}. This is since important parts of the work on prediction in the probabilistic scenario include some assumption on the stationarity of the measure governing the
process, such as stationarity or asymptotically mean stationarity
\cite{Gray_Kieffer80}.\footnote{An important exception is the Kalman
filter \cite[Section 7.7]{Porat94}.} In the scandiction scenario, however, the properties of the output sequence are not easy to determine, and it is possible, in general, that the output sequence is not stationary or ergodic even if the input data array is. Thus, although under certain assumptions, one can use a single universal predictor, applied to any scan in a certain set of scans, this is not the case in general. 
\subsection{Universal Scandiction for Mixing Random Fields}\label{subsec. universality for mixing rfs}
The proof of Theorem \ref{theo. universal finite set scandictability}
established the universality of $(\hat{\Psi},\hat{F})_n$ under the
expected cumulative loss criterion. In order to establish its
universality in the $Q$-a.s.\ sense, we examine the conditions on the measure $Q$ such that the following equality holds. 
\begin{equation}\label{eq. what we need for Qas}
\lim_{n \to \infty} \frac{1}{K^2}\sum_{i=1}^{K^2}{L_j(x^i)} =
E_{Q_{V_m}}L_j(X^0) \quad Q-a.s.
\end{equation}
To this end, we briefly review the conditions for the individual ergodic theorem for general dynamical systems given in \cite{Tempelman72}, specialized for $\Z^2$. Let $\{A_n\}$ be a sequence of subsets of $\Z^2$. For each $n$, the set $A_n$ is the set of sites over which the arithmetical average is taken. Let $A \triangle B$ denote the symmetric difference between the sets $A$ and $B$, $A \cup B \smallsetminus A\cap B$, and remember that $\tau_i(x)_j=x_{j+i}$.
\par\noindent\textbf{Condition 1 (\cite[$E1'$]{Tempelman72}).} For all $i \in \Z^2$,
\begin{equation}
\lim_{n \to \infty} \frac{|A_n \triangle \tau_i(A_n)|}{|A_n|}=0.
\end{equation}
\par\noindent\textbf{Condition 2 (\cite[$E3''$]{Tempelman72}).} There exists a constant $C_1 < \infty$ such that for all $n$,
\begin{equation}
|k:k=i-j, \quad  i,j \in A_n| \leq C_1|A_n|.
\end{equation}
\par\noindent\textbf{Condition 3 (\cite[$E4$]{Tempelman72}).} There exists a sequence of measurable sets $\{M_n\}$ such that,
\begin{equation}
\liminf_{n \to \infty} \frac{|k: k=i+j, \quad  i \in A_n, j \in M_n|}{|M_n|}=C_2 < \infty.
\end{equation}
By \cite[Theorem 6.1$'$]{Tempelman72}, if the sequence $\{A_n\}$ satisfies conditions 1-3, then, for any stationary random field $X$ with $E|X_0| < \infty$, we have,
\begin{equation}
\lim_{n \to \infty} \frac{1}{|A_n|}\sum_{i \in A_n}X_i = E\{X_0 | \calI\} \quad Q-a.s.,
\end{equation}
where $Q$ is the measure governing $X$ and $\calI$ is the $\sigma$-algebra of invariant sets of $\Omega$, that is,
\begin{equation}
A \in \calI \text{ iff } \tau_i(A)=A \quad \text{for all } i \in \Z^2.
\end{equation}
If $Q$ is ergodic, namely, for each $A \in \calI$, $Q(A) \in \{0,1\}$, then $E\{X_0 | \calI\}$ is deterministic and equals $EX_0$. 

Clearly, since $L_j(x^i)$ depends on a set of $m^2$ sites, with the average in taken over the sets $A_n = \{i:i=m\cdot j, j\in V_K\}$, \eqref{eq. what we need for Qas} may not hold, even if $Q$ is ergodic, as, for example, Condition 1 is not satisfied.\footnote{In fact, Tempelman's work \cite{Tempelman72} also includes slightly weaker conditions, but neither are satisfied in the current setting.} These two obstacles can be removed by defining an alternative random field, $\tilde{X}$, over the set of sites $m\cdot \Z^2=\{j:j=m\cdot i, i \in \Z^2\}$, where $\tilde{X_i}$ equals $L_j(X^k)$ and $X^k$ is the corresponding $m \times m$ block of $X$. Note that since the loss function $l(\cdot)$ is bounded and $m$ is finite, $E|\tilde{X_0}| < \infty$. It is not hard to see that conditions 1-3 are now satisfied (with the new space being $m\cdot \Z^2$). However, for $E\{\tilde{X_0}|\calI_m\}$ to be deterministic, where $\calI_m$ is the $\sigma$-algebra of $m$-invariant sets,
\begin{equation}
A \in \calI_m \text{ iff } \tau_j(A)=A \quad \text{for all } j=i\cdot m, i \in \Z^2,
\end{equation}
it is required that $\calI_m$ is the trivial $\sigma$-algebra. In other words, block ergodicity of $Q$ is required.

We now show that if the measure $Q$ is strongly mixing, then it is block-ergodic for any finite block size. For $A,B \in \Z^2$, define
\begin{equation}
\alpha^Q(A,B)=\sup\{|Q(U \cap V)-Q(U)Q(V)|, U \in \sigma(X_A),V \in \sigma(X_B)\},
\end{equation}
where $\sigma(X_B)$ is the smallest sigma algebra generated by $X_B$. Let $\alpha^Q_{a,b}(k)$ denote the strong mixing coefficient
\cite[Sec. 1.7]{Guyon95} of the random field $Q$
\begin{equation}
\alpha^Q_{a,b}(k)=\sup\{\alpha^Q(A,B), |A| \leq a, |B| \leq b, d(A,B)
\geq k\},
\end{equation}
where $d$ is a metric on $\Z^2$ and $d(A,B)$ is the distance between the
closest points, i.e., $d(A,B)=\min_{i\in A,j\in B}d(i,j)$. Assume now that $Q$
is strongly mixing in the sense that for all $a,b \in \N \cup \{\infty\}$,
$\alpha^Q_{a,b}(k) \to 0$ as $k \to \infty$. It is easy to see that $Q(A)\in \{0,1\}$ for all $A \in \calI_m$. Indeed,
\begin{equation}
\lim_{d(i,0) \to \infty}|Q(\tau_{i\cdot m}(A) \cap A)-Q(\tau_{i\cdot
m}(A))Q(A)|=0,
\end{equation}
however, since $A$ is $m$-invariant, $\tau_{i\cdot m}(A)=A$ and thus $Q(A) = Q(A)^2$. Hence $Q$ is $m$-block ergodic for each $m$ (i.e., totally ergodic). 

The following theorem asserts that under the assumption that the
random field $Q$ is strongly mixing, the results of Theorem \ref{theo.
universal finite set scandictability} apply in the a.s. sense as well.
%
\begin{theorem}\label{theo. a.s. counterpart of finite set theo.}
Let $X$ be a stationary strongly mixing random field with a probability measure
$Q$. Let $\calF=\{\calF_n\}$ be a sequence of finite sets of scandictors and assume
that $U_{\calF}(l,Q)$ exists. Then, if the universal algorithm suggested in the
proof of Theorem \ref{theo. universal finite set scandictability} uses a fixed
block size $m$, we have
\begin{equation} \label{eq. inq. between the liminf - a.s}
\liminf_{n \rightarrow \infty}\frac{1}{|V_n|}L_{alg}(X_{V_n}) \leq
U_{\calF}(l,Q) + \delta(m) \quad Q-a.s.
\end{equation}
for any such $Q$ and some $\delta(m)$ such that $\delta(m) \to 0$ as $m \to
\infty$.
\end{theorem}
\begin{proof}
For each $x_{V_n}$, we have,
\begin{eqnarray}
\frac{1}{|V_n|}L_{min}(x_{V_n}) &=& \frac{1}{|V_n|}\min_{(\Psi,F)_j \in
\calF_{m}} \sum_{i=1}^{(K+1)^2}{L_j(x^i)}
\nonumber\\
& \leq & \frac{1}{|V_n|} \left(\min_{(\Psi,F)_j \in \calF_{m}}
\sum_{i=1}^{K^2}{L_j(x^i)} +2Km(n-Km)l_{max} + (n-Km)^2l_{max} \right)
\nonumber\\
& \leq & \frac{1}{|V_m|} \min_{(\Psi,F)_j \in \calF_{m}}
\frac{1}{K^2}\sum_{i=1}^{K^2}{L_j(x^i)} + 2\frac{m}{n}l_{max}.
\end{eqnarray}
By Proposition \ref{prop. compete with a set of
  scan. block-wise},
\begin{equation}
\frac{1}{|V_n|}\bar{L}_{alg}(x_{V_n}) \leq \frac{1}{|V_n|}L_{min}(x_{V_n}) +
\frac{m(n+m)}{|V_n|}\sqrt{\log
    \lambda}\frac{l_{max}}{\sqrt{2}}.
\end{equation}
Thus,
\begin{eqnarray}
\liminf_{n \to \infty}\frac{1}{|V_n|}\bar{L}_{alg}(x_{V_n}) &\leq&  \liminf_{n \to \infty}\frac{1}{|V_n|}L_{min}(x_{V_n})
\nonumber\\
&\leq& \frac{1}{|V_m|} \liminf_{n \to \infty}\min_{(\Psi,F)_j \in \calF_{m}}
\frac{1}{K^2}\sum_{i=1}^{K^2}{L_j(x^i)}
\nonumber\\
& \leq &\frac{1}{|V_m|} \min_{(\Psi,F)_j \in \calF_{m}}\liminf_{n \to \infty}
\frac{1}{K^2}\sum_{i=1}^{K^2}{L_j(x^i)}.
\end{eqnarray}
Since $K \to \infty$ as $n \to \infty$, by the block ergodicity of $Q$ and the
fact that for finite $m$ and each $(\Psi,F)_j \in \calF_{m}$, $L_j(X)$ is a
bounded function, it follows that
\begin{equation}
\liminf_{n \to \infty} \frac{1}{K^2}\sum_{i=1}^{K^2}{L_j(x^i)} =
E_{Q_{V_m}}L_j(X^0) \quad Q-a.s.
\end{equation}
Finally, since $U_{\calF}(l,Q)$ exists, there exists $\delta(m)$ such that
$\delta(m) \to 0$ as $m \to \infty$ and we have
\begin{equation}
\liminf_{n \to \infty}\frac{1}{|V_n|}\bar{L}_{alg}(x_{V_n}) \leq
U_{\calF}(l,Q)+\delta(m)\quad Q-a.s.
\end{equation}
The fact that $\bar{L}_{alg}(x_{V_n})$ converges to $L_{alg}(x_{V_n})$
a.s. is clear from the proof of Proposition \ref{prop. compete with a
  set of scan. block-wise, a.s.}.
\end{proof}
Very similar to Theorem \ref{theo. a.s. counterpart of finite set theo.}, we also have the following corollary.
%
\begin{corollary}\label{cor. a.s. performance for achieving U}
Let $X$ be a stationary strongly mixing random field over a finite alphabet $A$ and a probability measure $Q$. Let the prediction space $D$ be either finite or bounded (with $l(x,F)$ then being Lipschitz in its second argument). Then, there
exists a sequence of scandictors
$\{(\Psi,F)_n\}$, independent of $Q$, for which
\begin{equation}
\liminf_{n \rightarrow \infty}\frac{1}{|V_n|}L_{alg}(X_{V_n}) \leq U(l,Q) +
\delta(m) \quad Q-a.s.
\end{equation}
for any such $Q$ and some $\delta(m)$ such that $\delta(m) \to 0$ as $m
\to \infty$. Thus, when $m \to \infty$, the performance of $\{(\Psi,F)_n\}$ equals the scandictability of the source, $Q-a.s.$
\end{corollary}
%

\subsection{Universal Scandiction for Individual Images}\label{subsec. universality for ind. images}
The proofs of Theorems \ref{theo. universal finite set scandictability}
and \ref{theo. a.s. counterpart of finite set theo.} relied on
the stationarity, or the stationarity and mixing property, of the random field $X$
(respectively). In the proof of Theorem \ref{theo. universal finite set scandictability}, we used the fact that the cumulative loss of
any scandictor $(\Psi,F)$ on a given block of data has the same
expected value as that on any other block. In the proof of Theorem
\ref{theo. a.s. counterpart of finite set theo.}, on the other hand,
the fact that the
Cesaro mean of the losses on finite blocks converges to a single
value, the expected cumulative loss, was used.

When $x$ is an individual image, however, the cumulative loss of the suggested
algorithm may be higher than that of the best scandictor in the scandictors set
since restarting a scandictor at the beginning of each block may result in
arbitrarily larger loss compared to the cumulative loss when the scandictor
scans the entire data. Compared to the prediction problem, in the scandiction scenario, if
the scanner is arbitrary, then different starting conditions may yield
different scans (i.e., a different reordering of the data) and thus arbitrarily
different cumulative loss, even if the predictor attached to it is very simple,
e.g., a Markov predictor. It is expected, however, that when the scandictors
have some structure, it will be possible to compete with finite sets of
scandictors in the individual image scenario.

In this subsection, we suggest a basic
scenario under which universal scandiction of individual images is possible.
Further research in this area is required, though, in order to identify larger
sets of scandictors under which universality is achievable. As mentioned earlier, since the exponential weighting algorithm used in the proofs of Theorems \ref{theo. universal finite set scandictability}
and \ref{theo. a.s. counterpart of finite set theo.} applied only \emph{block-wise} scandictors, i.e., scandictors which scan every block of the data separately from all other blocks, stationarity or stationarity and ergodicity \emph{of the data} were required in order to prove its convergence. Here, since the data is an individual image, we impose restrictions on the families of scandictors in order to achieve meaningful results (this reasoning is analogous to that described in \cite[Section I-B]{Mer_Fed98} for the prediction problem). The first restriction is that the scanners with which we compete are such that the actual path taken by each scanner when it is applied in a block-wise order has some kind of an overlap (in a sense which will be defined later) with the path taken when it is applied to the whole image. The second restriction is that the predictors are Markovian of finite order (i.e., the prediction depends only on the last $k$ symbols seen, for some finite $k$). Note that the first restriction does not restrict us to compete only with scandictors which operate in a block-wise order, only requires that the excess loss induced when the scandictors operate in a block-wise order, compared to operating on the entire image, is not too large, if, in addition, the predictor is Markovian. 

The following definition, and the results which follow, make the above requirements precise. For two scanners $\Psi$ and $\Psi'$ for the data array $x_B$, define
$N_{B,K}(x_B,\Psi,\Psi')$ as the number of sites in $B$ such that their
immediate past (context of unit length) under $\Psi$ is contained in the
context of length $K$ under $\Psi'$, namely,
\begin{equation}
N_{B,K}(x_B,\Psi,\Psi') = \left|\left\{1 \leq i \leq  |B| : \exists_{1 \leq j \leq |B|,k\leq K} \quad (\Psi_i,\Psi_{i-1})=(\Psi'_j,\Psi'_{j-k}) \right\}\right|.
\end{equation}
Note that in the above definition, a ``context" of size $w$ for a site in $B$
refers to the \emph{set of $w$ sites} which precede it in the discussed scan,
and not their actual values. When $\{\Psi_n\}$ is a sequence of scanners, where
$\Psi_n$ is a scanner for $V_n$, it will be interesting to consider the number
of sites in $B \subset V_{n_2}$, where $B$ is an $n_1 \times n_1$ rectangle,
$n_1 \leq n_2$, such that their immediate past under $\Psi_{n_2}$ (applied to
$V_{n_2}$) is contained in the  context of length $K$ under $\Psi_{n_1}$
(applied to $B$), that is
\begin{equation}
N_{B,K}(x_{B},\Psi_{n_2},\Psi_{n_1}) = \left|\left\{1 \leq i \leq |B| : \exists_{1 \leq j \leq |V_{n_2}|,k\leq K} \quad (\Psi_{n_2,i},\Psi_{n_2,i-1})=(\Psi_{n_1,j},\Psi_{n_1,j-k}) \right\}\right|,
\end{equation}
where $\Psi_{n_\cdot,i}$ is the $i$'th site the scanner $\Psi_{n_\cdot}$ visits. The following proposition is proved in Appendix \ref{app. proof of prop. two identical scanners on B}.
%
\begin{proposition}\label{prop. two identical scanners on B}
Consider two scanners $\Psi$ and $\Psi'$ for $B$ such that for any individual image $x_B$ we have
\begin{equation}\label{eq. cond. for identical scans}
\frac{N_{B,K}(x_B,\Psi,\Psi')}{|B|} = 1 - o(|B|).
\end{equation}
Then, for any $x_B$,
\begin{equation}
L_{(\Psi',F^{Kw,opt})}(x_B) \leq L_{(\Psi,F^{w,opt})}(x_B) + o(|B|)(K+1)^{w-1} l_{max},
\end{equation}
where for each scandictor $(\Psi, F^{w,opt})$, $F^{w,opt}$ denotes the optimal
$w$-order Markov predictor for the scan $\Psi$.
\end{proposition}
Note that in order to satisfy the condition in \eqref{eq. cond. for identical
scans} for \emph{any} array $x_B$, it is likely (but not a compulsory) that
both $\Psi$ and $\Psi'$ are data-independent scans. However, they need not be
identical. If, for example, $\Psi$ is a raster scan from left to right, and
$\Psi'$ applies the same left to right scan, but with a different ordering of
the rows, then the condition is satisfied for any $x_B$.

The result of Proposition \ref{prop. two identical scanners on B} yields the
following corollary, which gives sufficient conditions on the scandictors sets
under which a universal scandictor for any individual image exists. The proof can be found in Appendix \ref{app. proof of cor. individual images 1}.
\begin{corollary}\label{cor. individual images 1}
Let $\{\calF_n\}$, $|\calF_n|=\lambda < \infty$, be a sequence of scandictor
sets, where \newline
$\calF_n=\{(\Psi^1_n,F^1),(\Psi^2_n,F^2),\ldots,(\Psi^\lambda_n,F^\lambda)\}$
is a set of scandictors for $V_n$. Assume that the predictors are Markov of
finite order $w$, the prediction space $D$ is finite, and that there exists
$m(n)=o(n)$ (yet $m(n) \to \infty$ as $n \to \infty$) such that for all $1 \leq
i \leq \lambda$, $n$, and $x_{V_n}$ we have
\begin{equation}\label{cond. for cor. individual images 1}
\frac{N_{B_{m(n)},K}\left(x_{B_{m(n)}},\Psi^i_n,\Psi^i_{m(n)}\right)}{m(n)^2} =
1 - o\left(m(n)^2\right),
\end{equation}
where $B_{m(n)}$ is any one of the $\Floor[\frac{n}{m(n)}]^2$ sub-blocks of
size $m(n)\times m(n)$ of $V_n$. Then, there exists a sequence of scandictors
$\{(\hat{\Psi},\hat{F})_n\}$ such that for any image $x$
\begin{equation}\label{eq. inq. between the
liminf ind.}
\liminf_{n \rightarrow \infty}
E\frac{1}{|V_n|}L_{(\hat{\Psi},\hat{F})_n}(x_{V_n}) \leq
\liminf_{n \rightarrow \infty}\min_{(\Psi,F) \in \calF_n}
\frac{1}{|V_n|}L_{(\Psi,F)}(x_{V_n})
\end{equation}
where the expectation in the l.h.s. of \eqref{eq. inq. between the
liminf ind.} is due to the possible randomization in $(\hat{\Psi},\hat{F})_n$.
\end{corollary}
Although the condition in \eqref{cond. for cor. individual images 1} is limiting, and may not be met by many data-dependent scans, Corollary \ref{cor. individual images 1} still answers on the affirmative the following basic question: do there exist scandictor sets for which one can find a universal scandictor in the individual image scenario? For example, by  Corollary \ref{cor. individual images 1}, if the scandictor set includes all raster-type scans (e.g., left-to-right, right-to-left, up-down, down-up, diagonal, etc.), accompanied with Markov predictors of finite order, then there exists a universal scandictor whose asymptotic normalized cumulative loss is less or equal than that of the best scandictor in the set, for any individual image $x$. The condition in \eqref{cond. for cor. individual images 1} is also satisfied for some well-known ``self-similar" space filling curves, such as the Sierpinski or Lebesgue curves \cite{Sagan94}.

%% file: sensitivity.tex
\section{Bounds on the Excess Scandiction Loss for Non-Optimal Scanners}\label{sec. Bounds}
While the results of Section \ref{sec. Universal Scandiction} establish the existence of a universal scandictor for all stationary random fields and bounded loss function (under the terms of Theorem \ref{theo. existence of an alg achieving U}), it is interesting to investigate, from both practical and theoretical reasons, what is the excess scandiction loss when non-optimal scanners are used. I.e., in this section we answer the following question: Suppose that, for practical reasons for example, one uses a non-optimal scanner, accompanied with the optimal predictor for that scan. How large is the excess loss incurred by this scheme with respect to optimal scandiction?

For the sake of simplicity,  we consider the scenario of predicting the next outcome
of a binary source, with $D = [0,1]$ as the prediction space.
Hence, $l:\{0,1\}\times[0,1] \to \R$ is the loss function. Furthermore, we assume deterministic scanner (though data-dependent, of course). The generalization to randomized scanners is cumbersome but straightforward.

Let $\phi_l$ denote the
Bayes envelope associated with $l$, i.e.,
\begin{equation}\label{def. phi - bayes env}
\phi_l(p)=\min_{q\in[0,1]}[(1-p)l(0,q)+pl(1,q)].
\end{equation}
We further define
\begin{equation}\label{def. epsilon_l}
\epsilon_l=\min_{\alpha,\beta}\max_{0\leq p\leq 1}|\alpha h_b(p)+\beta
- \phi_l(p)|,
\end{equation}
where $h_b(p)$ is the binary entropy function. Thus $\epsilon_l$ is the error in approximating $\phi_l(p)$ by the best affine function of $h_b(p)$. For example, when $l$ is the Hamming loss function, denoted by $l_H$, we have $\epsilon_{l_H}=0.08$ and when $l$ is the squared error, denoted by $l_s$, $\epsilon_{l_s}=0.0137$. For the log loss, however, the expected instantaneous loss equals the conditional entropy, hence the expected cumulative loss coincides with the entropy, which is invariant to the scan, and we have $\epsilon_l=0$. To wit, the scan is inconsequential under log loss.

Although the definitions of $\phi_l(p)$ and $\epsilon_l$ refer to the binary scenario, the results below (Theorem \ref{theo. upper bound on the redundancy for non-optimal scanner} and Propositions \ref{prop. abs diff entropy and loss} and \ref{prop. abs diff entropy and loss with scanning}) hold for larger alphabets, with $\epsilon_l$ defined as in \eqref{def. epsilon_l}, with the maximum ranging over the simplex of all distributions on the alphabet, and $h(p)$ (replacing $h_b(p)$) and $\phi_l(p)$ denoting the entropy and Bayes envelope of the distribution $p$, respectively.

Let $\Psi$ be any (possibly data dependent) scan, and let
$E_{Q_B}\frac{1}{|B|}L_{(\Psi,F^{opt})}(X_B)$ denote the expected normalized cumulative
loss in scandicting $X_B$ with the scan $\Psi$ and the optimal
predictor for that scan, under the loss function $l$.  Remembering that $U(l,Q_B)$ denotes the scandictability of $X_B$ w.r.t the
loss function $l$, namely,
\mbox{$U(l,Q_B)=\inf_{\Psi}E_{Q_B}\frac{1}{|B|}L_{(\Psi,F^{opt})}(X_B)$}, our main result in this section is the following.
%
\begin{theorem}\label{theo. upper bound on the redundancy for non-optimal scanner}
Let $X_B$ be an arbitrarily distributed binary field. Then, for
any scan $\Psi$,
\begin{equation}\label{eq. upper bound on redundancy for non-optimal scanner}
\left|E_{Q_B}\frac{1}{|B|}L_{(\Psi,F^{opt})}(X_B)-U(l,Q_B)\right| \leq 2\epsilon_l.
\end{equation}
\end{theorem}
That is, the excess loss incurred by applying \emph{any scanner} $\Psi$, accompanied with the optimal predictor for that scan, with respect to optimal scandiction is not larger than $2\epsilon_l$. 

To prove Theorem \ref{theo. upper bound on the redundancy for non-optimal scanner}, we first introduce a prediction result (i.e., with no data reordering) on the error in estimating the cumulative loss of a predictor under a loss function $l$ with the best affine function of the entropy. We then generalize this result to the multi-dimensional case.  
%
\begin{proposition}\label{prop. abs diff entropy and loss}
Let $X^n$ be an arbitrarily distributed binary $n$-tuple and let
$EL_l^{opt}(X^n)$ denote the expected cumulative loss in
predicting $X^n$ with the optimal distribution-dependent scheme for the
loss function $l$. Then,
\begin{equation}
\left|\alpha_l \frac{1}{n}H(X^n)+\beta_l-\frac{1}{n}EL_l^{opt}(X^n)\right| \leq \epsilon_l,
\end{equation}
where $\alpha_l$ and $\beta_l$ are the achievers of the minimum in
\eqref{def. epsilon_l}.
\end{proposition}
\begin{proof}
Let $\alpha_l$ and $\beta_l$ be the achievers of the minimum in
\eqref{def. epsilon_l}. We have,
\begin{eqnarray}
\left| \alpha_l\frac{1}{n}H(X^n)+\beta_l-\frac{1}{n}EL_l^{opt}(X^n) \right|
\nonumber\\
&&\hspace{-4.5cm}= \Bigg|
  \frac{1}{n}\sum_{t=1}^{n}\sum_{x^t} P(x^{t-1})
\nonumber\\
&&\hspace{-3.5cm}
  \left[ -\alpha_l P(x_t|x^{t-1})\log
      P(x_t|x^{t-1}) + P(x_t|x^{t-1})\beta_l -
      P(x_t|x^{t-1})l(x_t,F_t^{opt}(x^{t-1}))\right] \Bigg|
\nonumber\\
&&\hspace{-4.5cm}\stackrel{(a)}{=} \left|
  \frac{1}{n}\sum_{t=1}^{n}{\sum_{x^{t-1}}{ P(x^{t-1}) \left[\alpha_l h_b(P(\cdot|x^{t-1}))+\beta_l-\phi_l(P(\cdot|x^{t-1})) \right] }} \right|
\nonumber\\
&&\hspace{-4.5cm}\leq
  \frac{1}{n}\sum_{t=1}^{n}{\sum_{x^{t-1}}{ P(x^{t-1}) \left|\alpha_l h_b(P(\cdot|x^{t-1}))+\beta_l-\phi_l(P(\cdot|x^{t-1})) \right| }}
\nonumber\\
&&\hspace{-4.5cm}\leq \frac{1}{n}\sum_{t=1}^{n}{\sum_{x^{t-1}}{
    P(x^{t-1}) \max_{p} \left|\alpha_l
      h_b(p)+\beta_l-\phi_l(p) \right| }}
\nonumber\\
&&\hspace{-4.5cm}= \max_{p} \left|\alpha_l
      h_b(p)+\beta_l-\phi_l(p) \right|
\nonumber\\
&&\hspace{-4.5cm}=\epsilon_l,
\end{eqnarray}
where $(a)$ is by the definition of $\phi_l(\cdot)$ and the optimality
of $F_t^{opt}$ with respect to $l$.
\end{proof}
The following proposition is the generalization of Proposition \ref{prop. abs diff entropy and loss} to the multi-dimensional case.
%
\begin{proposition}\label{prop. abs diff entropy and loss with scanning}
Let $X_B$ be an arbitrarily distributed binary random field. Then, for
any scan $\Psi$,
\begin{equation}\label{eq. bound of prop. abs diff with scanning}
\left|\alpha_l \frac{1}{|B|}H(X_B)+\beta_l-E_{Q_B}\frac{1}{|B|}L_{(\Psi,F^{opt})}(X_B)\right| \leq \epsilon_l,
\end{equation}
where $\alpha_l$ and $\beta_l$ are the achievers of the minimum in
\eqref{def. epsilon_l}.
\end{proposition}
For data-independent scans, the proof follows the proof of Proposition \ref{prop. abs diff entropy and loss} verbatim by applying it to the reordered
$|B|$-tuple $X_{\Psi_1},\ldots,X_{\Psi_{|B|}}$ and remembering that
$H(X_B)=H(X_{\Psi_1},\ldots,X_{\Psi_{|B|}})$. For data-dependent scans, the
proof is similar, but requires more caution.
\begin{proof}[Proof of Proposition \ref{prop. abs diff entropy and loss with scanning}.]
Let $\alpha_l$ and $\beta_l$ be the achievers of the minimum in
\eqref{def. epsilon_l}. For a given data array $x_B$,
$\Psi_1,\Psi_2(x_{\Psi_1}),\ldots,\Psi_{|B|}(x^{\Psi_{|B|-1}})$ are fixed, and
merely reflect a reordering of $x_B$ as a $|B|$-tuple. Thus,
\begin{eqnarray}
\left| \alpha_l\frac{1}{|B|}H(X_B)+\beta_l-E_{Q_B}\frac{1}{|B|}L_{(\Psi,F^{opt})}(X_B) \right|
\nonumber\\
&&\hspace{-7cm}=
\left|
  \frac{1}{|B|}\sum_{x_B}{\left( -\alpha_l P(x_B)\log
      P(x_B)  -
      P(x_B)\sum_{t=1}^{|B|}{l(x_{\Psi_t},F_t^{opt}(x^{\Psi_{t-1}}))}\right) + \beta_l} \right|
\nonumber\\
&&\hspace{-7cm}=\left|
  \frac{1}{|B|}\sum_{x_B}{\left( -\alpha_l P(x_B)\sum_{t=1}^{|B|}{\log P(x_{\Psi_t}|x^{\Psi_{t-1}})} -
      P(x_B)\sum_{t=1}^{|B|}{l(x_{\Psi_t},F_t^{opt}(x^{\Psi_{t-1}}))}\right)+ \beta_l  } \right|
\nonumber\\
&&\hspace{-7cm}=\left|
  \frac{1}{|B|}\sum_{t=1}^{|B|}\sum_{x_B} P(x_B) \Big( -\alpha_l \log P(x_{\Psi_t}|x^{\Psi_{t-1}}) + \beta_l -
      l(x_{\Psi_t},F_t^{opt}(x^{\Psi_{t-1}})) \Big)
    \right|. \label{eq. first half of the proof of claim 2}
\end{eqnarray}
Fix $t=t_0$ in the sum over $t$. Consider all data arrays $x_B$ such that for a specific scanner $\Psi$ we have 
\begin{equation}
\left\{\Psi_1,\Psi_2(x_{\Psi_1}),\ldots,\Psi_{t_0-1}(x_{\Psi_1},\ldots,x_{\Psi_{t_0-2}})\right\}=I(\Psi),
\end{equation}
where $I(\Psi) \subset B$ is a fixed set
of sites, and $(x_{\Psi_1},\ldots,x_{\Psi_{t_0-1}})=\underbar{a}$, for some
$\underbar{a}\in\{0,1\}^{t_0-1}$. In this case, $\Psi_t(x^{\Psi_{t_0-1}})$ is
also fixed, and since the term in the parentheses of \eqref{eq. first
half of the proof of claim 2} depends only on $I$, $\underbar{a}$ and
$x_{\Psi_{t_0}}$, we have
\begin{multline}
\sum_{x^n} P(x^n) \Big( -\alpha_l \log P(x_{\Psi_{t_0}}|x^{\Psi_{{t_0}-1}}) + \beta_l -
      l(x_{\Psi_{t_0}},F_{t_0}^{opt}(x^{\Psi_{{t_0}-1}})) \Big)
\\ = \sum_{I,\underbar{a}} P(x_I=\underbar{a})
\sum_{x_{\Psi_{t_0}}\in \{0,1\}}
P(x_{\Psi_{t_0}}|x_I=\underbar{a}) \Big( -\alpha_l \log P(x_{\Psi_{t_0}}|x_I=\underbar{a}) + \beta_l -
      l(x_{\Psi_{t_0}},F_{t_0}^{opt}(\underbar{a})) \Big).
\end{multline}
Consequently,
\begin{eqnarray}
\left| \alpha_l\frac{1}{|B|}H(X_B)+\beta_l-E_{Q_B}\frac{1}{|B|}L_{(\Psi,F^{opt})}(X_B) \right|
\nonumber\\
&&\hspace{-7cm} \leq
\frac{1}{|B|}\sum_{t=1}^{|B|}\sum_{I,\underbar{a}}
P(x_I=\underbar{a})
\nonumber\\
&&\hspace{-6.5cm}
\left| \sum_{x_{\Psi_t}\in
    \{0,1\}} P(x_{\Psi_t}|x_I=\underbar{a}) \Big( -\alpha_l \log P(x_{\Psi_t}|x_I=\underbar{a}) + \beta_l -
      l(x_{\Psi_t},F_t^{opt}(\underbar{a})) \Big) \right|
\nonumber\\
&&\hspace{-7cm} =
\frac{1}{|B|}\sum_{t=1}^{|B|}\sum_{I,\underbar{a}}
P(x_I=\underbar{a}) \left|\alpha_l h_b(P(\cdot|x_I=\underbar{a}))+\beta_l-\phi_l(P(\cdot|x_I=\underbar{a})) \right|
\nonumber\\
&&\hspace{-7cm}\leq
\frac{1}{|B|}\sum_{t=1}^{|B|} \sum_{I,\underbar{a}}
P(x_I=\underbar{a})  \max_{p} \left|\alpha_l
      h_b(p)+\beta_l-\phi_l(p) \right|
\nonumber\\
&&\hspace{-7cm}=
\max_{p} \left|\alpha_l
      h_b(p)+\beta_l-\phi_l(p) \right|
\nonumber\\
&&\hspace{-7cm}=\epsilon_l.
\end{eqnarray}
\end{proof}
It is now easy to see why Theorem \ref{theo. upper bound on the redundancy for non-optimal scanner} holds.
%
\begin{proof}[Proof of Theorem \ref{theo. upper bound on the redundancy for non-optimal scanner}.]
The proof is a direct application of Proposition \ref{prop. abs diff entropy and loss with scanning}, as for any scan $\Psi$,
\begin{eqnarray}
\left|E_{Q_B}\frac{1}{|B|}L_{(\Psi,F^{opt})}(X_B)-U(l,Q_B)\right| 
\nonumber\\&&\hspace{-5cm}\leq
\bigg|\frac{\alpha_l}{|B|}H(X_B)+\beta_l-E_{Q_B}\frac{1}{|B|}L_{(\Psi,F^{opt})}(X_B)\bigg|+\bigg|\frac{\alpha_l}{|B|}H(X_B)+\beta_l-U(l,Q_B)\bigg|
\nonumber\\
&&\hspace{-5cm}\leq 2\epsilon_l.
\end{eqnarray}
\end{proof}
At this point, a few remarks are in order. For the bound in Theorem \ref{theo. upper bound on the redundancy for non-optimal scanner} to be tight, the following conditions should be met. First, equality is required in \eqref{eq. bound of prop. abs diff with scanning} for both the scan $\Psi$ and the optimal scan (which achieves $U(l,Q_B)$). It is not hard to see that for a given scan $\Psi$, equality in \eqref{eq. bound of prop. abs diff with scanning} is achieved if and only if $P(\cdot|x^{\Psi_{t-1}}) = p$ for all $x^{\Psi_{t-1}}$, where $p$ is a maximizer of \eqref{def. epsilon_l}. However, for \eqref{eq. upper bound on redundancy for non-optimal scanner} to be tight, it is also required that
\begin{equation}
\frac{\alpha_l}{|B|}H(X_B)+\beta_l-E_{Q_B}\frac{1}{|B|}L_{(\Psi,F^{opt})}(X_B)=-\frac{\alpha_l}{|B|}H(X_B)-\beta_l+U(l,Q_B),
\end{equation}
so the triangle inequality is held with equality. Namely, it is required that under the scan $\Psi$, for example, $P(\cdot|x^{\Psi_{t-1}}) = p$ for all $x^{\Psi_{t-1}}$, where $p$ is such that $\alpha_l h_b(p) + \beta_l -\phi_l(p) = \epsilon_l$, yet under the optimal scan, say $\Psi'$, $P(\cdot|x^{\Psi'_{t-1}}) = p'$ for all $x^{\Psi'_{t-1}}$, where $p'$ is such that $\alpha_l h_b(p') + \beta_l -\phi_l(p') = -\epsilon_l$. Clearly this is not always the case, and thus, generally, the bound in Theorem \ref{theo. upper bound on the redundancy for non-optimal scanner} is not tight. Indeed, although under a different setting (individual images), in subsection \ref{subsec. ind. and PH} we derive a tighter upper bound on the excess loss for the specific case of Hamming loss. Using this bound, it is easy to see that the $0.16$ bound given here (as $\epsilon_l=0.08$ for Hamming loss) is only a worst case, and typically much tighter bounds on the excess loss apply, depending on the image compressibility. For example, consider a $1$st order symmetric Markov chain with transition probability $1/4$. Scanning this source in the trivial (sequential) order results in an error rate of $1/4$. By \cite{Mer_Weiss03}, this is indeed the optimal scanning order for this source, as it can be represented as an autoregressive process whose innovation process has a maximum entropy distribution with respect to the Hamming distance. The ``odds-then-evens" scan\footnote{An ``odds-then-evens" scanner for a one-dimensional vector $x_1^n$, first scans all the sites with an odd index, in an ascending order, then all the sites with an even index.}, however, which was proved useful for this source but with larger transition probabilities (larger than $1/2$, \cite{Mer_Weiss03}), results in an error rate of $5/16$, which is $1/16$ away from the optimum. It is not hard to show that different transition probabilities result in lower excess loss. 
\subsection{Individual Images and the Peano-Hilbert Scan}\label{subsec. ind. and PH}
In this subsection, we seek analogous results for the individual image scenario. Namely, the data array $x_B$ has no stochastic model. A scandictor $(\Psi,F)$, in this case, wishes to minimize the cumulative loss over $x_{V_n}$, that is, $L_{(\Psi,F)}(x_{V_n})$ as defined in \eqref{eq. cumulative loss}. 

In this setting, although one can easily define an empirical probability measure, the invariance of the entropy $H(X^n)$ to the reordering of the components, which stood at the heart of Theorem \ref{theo. upper bound on the redundancy for non-optimal scanner}, does not hold for \emph{any} reordering (scan) and any finite $n$. Thus, we limit the possible set of scanners to that of the finite state machines discussed earlier. Moreover, in the sequel, we do not bound the difference in the scandiction losses of any two scandictors from that set, only that between the Peano-Hilbert scan (which is asymptotically optimal for compression of individual images \cite{Lempel_Ziv86}) and any other finite state scanner (both accompanied with an optimal Markov predictor), or between two scans (finite state or not) for which the FS compressibility of the resulting sequence is the same.  

We start with several definitions. Let $\Psi_{B}$ be a scanner for the data array $x_B$. Let $x_1^{|B|}$ be the sequence resulting from scanning $x_B$ with $\Psi_{B}$. Fix $k < |B|$ and for any $s \in \{0,1\}^{k+1}$ define the empirical distribution of order $k+1$ as
\begin{equation}
\hat{P}_{\Psi_{B}}^{k+1}(s) = \frac{1}{|B|-k}\left|\left\{ k < i \leq |B|: x_{i-k}^{i}=s \right\}\right|.
\end{equation}
The distributions of lower orders, and the conditional distribution are derived from $\hat{P}_{\Psi_{B}}^{k+1}(s)$, i.e., for $s' \in \{0,1\}^k$ and $x \in \{0,1\}$ we define
\begin{equation}
\hat{P}_{\Psi_{B}}^{k+1}(s')=\hat{P}_{\Psi_{B}}^{k+1}([s',0])+\hat{P}_{\Psi_{B}}^{k+1}([s',1])
\end{equation}
and
\begin{equation}
\hat{P}_{\Psi_{B}}^{k+1}(x|s') = \frac{\hat{P}_{\Psi_{B}}^{k+1}([s',x])}{\hat{P}_{\Psi_{B}}^{k+1}(s')},
\end{equation}
where $0/0$ is defined as $1/2$ and $[\cdot,\cdot]$ denotes string concatenation.\footnote{Note that defining $\hat{P}_{\Psi_{B}}^{k+1}(x|s')$, $s' \in \{0,1\}^k$ as $\frac{\hat{P}_{\Psi_{B}}^{k+1}([s',x])}{\hat{P}_{\Psi_{B}}^{k}(s')}$ is not consistent since generally $\hat{P}_{\Psi_{B}}^{k}(s') \ne \hat{P}_{\Psi_{B}}^{k+1}([s',0])+\hat{P}_{\Psi_{B}}^{k+1}([s',1])$.} Let $\hat{H}_{\Psi_{B}}^{k+1}(X|X^k)$ be the empirical conditional entropy of order $k$, i.e.,
 \begin{equation} 
\hat{H}_{\Psi_{B}}^{k+1}(X|X^k) = -\sum_{s \in \{0,1\}^k}{\hat{P}_{\Psi_{B}}^{k+1}(s)\sum_{x \in \{0,1\}}{\hat{P}_{\Psi_{B}}^{k+1}(x|s)\log \hat{P}_{\Psi_{B}}^{k+1}(x|s)}}.
\end{equation}
Finally, denote by $F^{k,opt}$ the optimal $k$-th order Markov predictor, in the sense that it minimizes the expected loss with respect to $\hat{P}_{\Psi_{B}}^{k+1}(\cdot|\cdot)$ and $x_1^{|B|}$. The following proposition is the individual image analogue of Proposition \ref{prop. abs diff entropy and loss with scanning}.
%
%
\begin{proposition}\label{prop. abs diff entropy and loss with scanning, ind. sequence}
Let $x_B$ be any data array. Let $\frac{1}{|B|}L_{(\Psi_{B},F^{k,opt})}(x_B)$ denote the normalized cumulative loss of the scandictor $(\Psi_{B}, F^{k,opt})$, where $\Psi_{B}$ is any (data dependent) scan and $F^{k,opt}$ is the optimal $k$-th order Markov predictor with respect to $\Psi_{B}$ and $l$. Then,
\begin{equation}
\left|\alpha_l \hat{H}_{\Psi_{B}}^{k+1}(X|X^k) +\beta_l-\frac{1}{|B|}L_{(\Psi_{B},F^{k,opt})}(x_B)\right| \leq \epsilon_l+\frac{k l_{max}}{|B|},
\end{equation}
where $\alpha_l$ and $\beta_l$ are the achievers of the minimum in
\eqref{def. epsilon_l}.
\end{proposition}
Since $x_{B}$ is an individual image, $x_1^{|B|}=\Psi_{B}(x_{B})$ is fixed. In that sense, the proof resembles that of Proposition \ref{prop. abs diff entropy and loss} and we write $x_t$ for the value of $x$ at the $t$-th site $\Psi_B$ visits. On the other hand, since the order of the predictor, $k$, is fixed, we can use $\hat{P}_{\Psi_{B}}^{k+1}(\cdot)$ and avoid the summation over the time index $t$. The complete details can be found in Appendix \ref{app. proof of prop. abs diff entropy and loss with scanning, ind. sequence}.

The bound in Proposition \ref{prop. abs diff entropy and loss with scanning, ind. sequence} differs from the one in Proposition \ref{prop. abs diff entropy and loss with scanning} for two reasons. First, it is only asymptotic due to the $O(k/|B|)$ term. Second, the empirical entropy $\hat{H}_{\Psi_{B}}^{k+1}(X|X^k)$ is not invariant to the scanning order. This is a profound difference between the random and the individual settings, and, in fact, is at the heart of \cite{Lempel_Ziv86}. In the random setting, the chain rule for entropies implies invariance of the entropy rate to the scanning order. This fact does not hold for a $k$-th order empirical distribution of an individual image, hence the usage of the Peano-Hilbert scanning order.\footnote{Yet, the Peano-Hilbert is by no means the \emph{only} optimal scan. We Elaborate on this issue later in this section.} Consequently, we cannot directly compare between any two scans. Nevertheless, Proposition \ref{prop. abs diff entropy and loss with scanning, ind. sequence} has the following two interesting applications, given by Proposition \ref{prop. diff PH other} and Corollary \ref{cor. diff two of same rho}.

For $\Psi=\{\Psi_n\}$, where $\Psi_n$ is a scan for $V_n$, and an infinite individual image $x$ define
\begin{equation}
L^k_{\Psi}(x) = \limsup_{n \to \infty}\frac{1}{|V_n|}L_{(\Psi_n,F^{k,opt})}(x_{V_n})
\end{equation}
and
\begin{equation} 
L_{\Psi}(x) = \lim_{k \to \infty}L^k_{\Psi}(x).
\end{equation}
Proposition \ref{prop. diff PH other} relates the asymptotic cumulative loss of any sequence of finite state scans $\Psi$ to that resulting from the Peano-Hilbert sequence of scans, establishing the Peano-Hilbert sequence as an advantageous scanning order for any loss function.  
\begin{proposition}\label{prop. diff PH other}
Let $x$ be any individual image. Let $PH$ denote the Peano-Hilbert sequence of scans. Then, for any sequence of finite state scans $\Psi$ and any loss function $l:\{0,1\}\times[0,1] \to \R$,
\begin{equation} 
L_{PH}(x) \leq L_{\Psi}(x) + 2 \epsilon_l.
\end{equation}
\end{proposition}
Before we prove Proposition \ref{prop. diff PH other}, define the asymptotic $k$-th order empirical conditional entropy under $\{\Psi_n\}$ as
\begin{equation}
\hat{H}^{k+1}_{\Psi}(x) = \limsup_{n \to \infty}\hat{H}_{\Psi_n}^{k+1}(X|X^k)
\end{equation}
and further define 
\begin{equation}
\hat{H}_{\Psi}(x) = \lim_{k \to \infty}\hat{H}^{k+1}_{\Psi}(x).
\label{def. hat H psi x}
\end{equation}
The existence of $\hat{H}_{\Psi}(x)$ is established later in the proof of Proposition \ref{prop. diff PH other}, where it is also shown that this limit equals $\lim_{k \to \infty}\limsup_{n \to \infty}\frac{1}{k}\hat{H}_{\Psi_n}^{k}(X^k)$. By \cite[Theorem 3]{Ziv_Lemp78}, the latter limit is no other than the asymptotic finite state compressibility of $x$ under the sequence of scans $\Psi$, namely,
\begin{eqnarray}
\lim_{k \to \infty}\limsup_{n \to \infty}\frac{1}{k}\hat{H}_{\Psi_n}^{k}(X^k) &=&
\rho(\Psi(x)) 
\nonumber\\
&=& \lim_{s \to \infty}\limsup_{n \to \infty}\rho_{E(s)}(\Psi_n(x_{V_n})),
\end{eqnarray}
where $\rho_{E(s)}(x_1^n)$ is the minimum compression ratio for $x_1^n$ over the class of all finite state encoders with at most $s$ states \cite[eq. (1)-(4)]{Ziv_Lemp78}. We may now introduce the following corollary.
\begin{corollary}\label{cor. diff two of same rho}
Let $\Psi_1$ and $\Psi_2$ be \emph{any} two sequences of scans such that $\hat{H}_{\Psi_1}(x)=\hat{H}_{\Psi_2}(x)$ (in particular, if both $\Psi_1$ and $\Psi_2$ are finite state sequences of scans they result in the same finite state compressibility). Then,
\begin{equation} 
\left|  L_{\Psi_1}(x) - L_{\Psi_2}(x) \right| \leq  2 \epsilon_l.
\end{equation}
for any loss function $l:\{0,1\}\times[0,1] \to \R$.
\end{corollary}
For a given sequence of scans $\Psi$, the set of scanning sequences $\Psi'$ satisfying $\hat{H}_{\Psi}(x)=\hat{H}_{\Psi'}(x)$ is larger than one might initially think. For example, a close look at the definition of finite state compressibility given in \cite{Ziv_Lemp78} shows that the finite state encoders defined therein allow \emph{limited scanning schemes}, as an encoder might read a large data set before its output for that data set is given. Thus, a legitimate finite state encoder in the sense of \cite{Ziv_Lemp78} may reorder the data in a block (of bounded length, as the number of states is bounded) before actually encoding it. Consequently, for any individual \emph{sequence} $x$ one can define several permutations having the same finite state compressibility. In the multidimensional scenario this sums up to saying that for each scanning sequence $\Psi$ there exist several different scanning sequences $\Psi'$ for which ${H}_{\Psi}(x)=\hat{H}_{\Psi'}(x)$.
\begin{proof}[Proof of Proposition \ref{prop. diff PH other}.]
For each $n$, $\Psi_n$ is a scanner for $V_n$. Thus, by Proposition \ref{prop. abs diff entropy and loss with scanning, ind. sequence}, we have
\begin{equation}
\left|\alpha_l \hat{H}_{\Psi_n}^{k+1}(X|X^k) +\beta_l-\frac{1}{|V_n|}L_{(\Psi_n,F^{k,opt})}(x_{V_n})\right| \leq \epsilon_l+\frac{k l_{max}}{|V_n|},
\end{equation}
Taking the limsup as $n \to \infty$ yields
\begin{equation}
\left|\alpha_l \limsup_{n \to \infty}\hat{H}_{\Psi_n}^{k+1}(X|X^k) +\beta_l-L^k_{\Psi}(x)\right| \leq \epsilon_l.
\end{equation}
For a stationary source, it is well known (e.g., \cite[Theorem 4.2.1]{Cov_Thom91}) that
$\lim_{k \to \infty} H(X_k|X_1^{k-1})$ exists and in fact
\begin{equation}
\lim_{k \to \infty} H(X_k|X_1^{k-1}) =\lim_{k \to \infty} \frac{1}{k}H(X_1^k).
\end{equation}
To this end, we show that the same holds for empirical entropies. We start by showing that $\limsup_{n \to \infty}\hat{H}_{\Psi_n}^{k+1}(X|X^k)$ is a decreasing sequence in $k$. Since conditioning reduces the entropy, it is clear that $\hat{H}_{\Psi_n}^{k+1}(X|X^k) \leq \hat{H}_{\Psi_n}^{k+1}(X|X^{k-1})$, where both are calculated using $\hat{P}^{k+1}_{\Psi_n}(\cdot)$. However, the above may not be true when $\hat{H}_{\Psi_n}^{k+1}(X|X^{k-1})$ is replaced by $\hat{H}_{\Psi_n}^{k}(X|X^{k-1})$, as the later is calculated using $\hat{P}^{k}_{\Psi_n}(\cdot)$. Nevertheless, using a simple counting argument, it is not too hard to show that for every $k$, $0 < j \leq k$ and $s \in \{0,1\}^i$, where $0 < i \leq j$, we have
\begin{equation}\label{eq. diff between Pk+1 and Pj}
\hat{P}^{k+1}_{\Psi_n}(s)-\frac{k+1-j}{|V_n|-k} \leq \hat{P}^{j}_{\Psi_n}(s) \leq \hat{P}^{k+1}_{\Psi_n}(s)+\frac{k+1-j}{|V_n|-k}.
\end{equation}
Thus, by the continuity of the entropy function, we have
\begin{eqnarray}
\limsup_{n \to \infty}\hat{H}_{\Psi_n}^{k+1}(X|X^k) &\leq& \limsup_{n \to \infty}\hat{H}_{\Psi_n}^{k+1}(X|X^{k-1})
\nonumber\\
&=& \limsup_{n \to \infty}\hat{H}_{\Psi_n}^{k}(X|X^{k-1}),
\end{eqnarray}
hence $\limsup_{n \to \infty}\hat{H}_{\Psi_n}^{k}(X|X^{k-1})$ is decreasing in $k$. Since it is a non negative sequence, $\hat{H}_{\Psi}(x)$ as defined in \eqref{def. hat H psi x} exists and we have
\begin{equation}\label{eq. bound on sensitivity for any scan, ind. image}
\left|\alpha_l \hat{H}_{\Psi}(x) +\beta_l-L_{\Psi}(x)\right| \leq \epsilon_l.
\end{equation}
We now show that indeed $\hat{H}_{\Psi}(x)$ equals $\rho(\Psi(x))$ for every sequence of finite state scans $\Psi$, hence when $\Psi$ is a sequence of finite state scans the results of \cite{Lempel_Ziv86} can be applied. The method is similar to that in \cite[Theorem 4.2.1]{Cov_Thom91}), with an adequate handling of empirical entropies. By \eqref{eq. diff between Pk+1 and Pj},
\begin{eqnarray}
\limsup_{n \to \infty}\frac{1}{k}\hat{H}^{k}_{\Psi_n}(X^k) &=& \limsup_{n \to \infty}\frac{1}{k}\sum_{i=1}^{k}\hat{H}^{k}_{\Psi_n}(X_i|X_1^{i-1})
\nonumber\\
&=&\limsup_{n \to \infty}\frac{1}{k}\sum_{i=1}^{k}\hat{H}^{i}_{\Psi_n}(X_i|X_1^{i-1}).
\end{eqnarray}
But the sequence $\limsup_{n \to \infty}\hat{H}^{i}_{\Psi_n}(X_i|X_1^{i-1})$ converges to $\hat{H}_{\Psi}(x)$ as $i \to \infty$, thus its Cesaro mean converges to the same limit and we have 
\begin{eqnarray}
\hat{H}_{\Psi}(x) &=& \lim_{k \to \infty}\limsup_{n \to \infty}\frac{1}{k}\hat{H}^{k}_{\Psi_n}(X^k)
\nonumber\\
&=& \rho(\Psi(x)).
\end{eqnarray}
Consider now the Peano-Hilbert sequence of finite state scans, denoted by $PH$. Let $\rho(x)$ denote the (finite state) compressibility of $x$ as defined in \cite[eq. (4)]{Lempel_Ziv86}. For any other sequence of finite state scans $\tilde{\Psi}$ we have
\begin{eqnarray}
\hat{H}_{PH}(x) &\leq& \rho(x)
\nonumber\\
&\leq& \hat{H}_{\tilde{\Psi}}(x),
\end{eqnarray}
where the first inequality is by \cite[eq. (9) and (16)]{Lempel_Ziv86} and the second is straightforward from the definition of $\rho(x)$. Finally,
\begin{eqnarray}
L_{PH}(x) &\stackrel{(a)}{\leq}& \epsilon_l + \beta_l + \alpha_l \hat{H}_{PH}(x)
\nonumber\\
&\leq &\epsilon_l + \beta_l + \alpha_l \hat{H}_{\tilde{\Psi}}(x)
\nonumber\\
&\stackrel{(b)}{\leq}& 2\epsilon_l +L_{\tilde{\Psi}}(x),
\end{eqnarray}
where $(a)$ and $(b)$ result from the application of \eqref{eq. bound on sensitivity for any scan, ind. image} to the sequences $PH$ and $\tilde{\Psi}$ respectively. 
\end{proof}
The proof of Corollary \ref{cor. diff two of same rho} is straightforward, using \eqref{eq. bound on sensitivity for any scan, ind. image} for both $\Psi_1$ and $\Psi_2$ and the triangle inequality.
%
\subsubsection{Hamming Loss} The bound in Proposition \ref{prop. diff PH other} is valid for any loss function $l:\{0,1\}\times[0,1] \to \R$. When $l$ is the Hamming loss, the resulting bound is 
\begin{equation}
L^{Hamming}_{PH}(x) \leq L^{Hamming}_{\Psi}(x) + 0.16,
\end{equation}
for any other finite state sequence of scans, namely, a uniform bound, regardless of the compressibility of $x$. However, using known bounds on the predictability of a sequence (under Hamming loss) in terms of its compressibility can yield a tighter bound.

\begin{figure}[ht]
\centering
\includegraphics[scale=0.55]{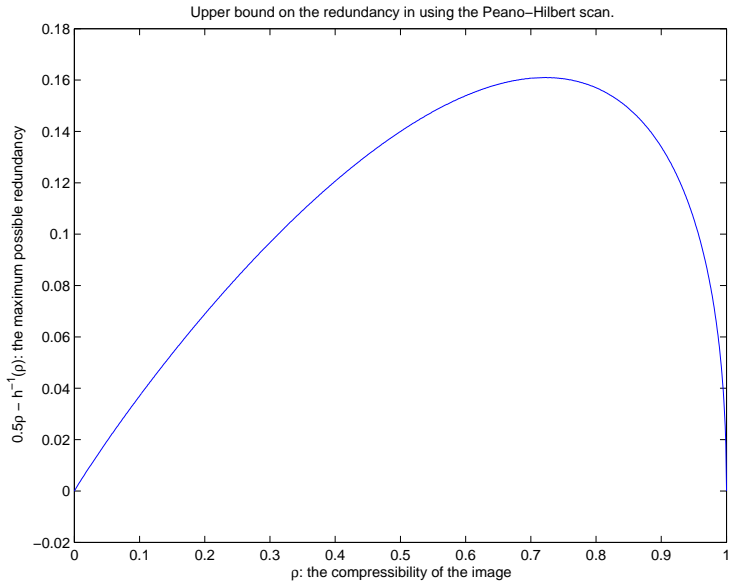}
\caption{A plot of $\frac{1}{2}\rho-h^{-1}(\rho)$. The maximum
  redundancy is not higher than $0.16$ in worst case, but will be much lower for more compressible arrays.}
\label{fig. redundancy}
\end{figure}

In \cite{Fed_Mer_Gut92}, Feder, Merhav and Gutman proved that for any
next-state function $g \in G_s$, where $G_s$ is the set of all possible
next state functions with $s$ states, and for any sequence $x_1^n$
\begin{eqnarray}\label{eq. bound on predic. in terms of compr.}
\mu(g,x_1^n) &\leq& \frac{1}{2}\rho(g,x_1^n),
\nonumber\\
\mu(g,x_1^n) &\geq& h_b^{-1}\left(\rho(g,x_1^n)\right),
\end{eqnarray}
where $\mu(g,\cdot)$ ($\rho(g,\cdot)$) is the best possible prediction (compression) performance when the next state function is $g$. Consequently, for \emph{any} two
finite-state scans $\Psi^1_n$ and $\Psi^2_n$ for $x_{V_n}$,
\begin{align}\label{eq. PH compared to Psi}
\min_{g \in G_s}\mu(g,&\Psi^1_n(x_{V_n}))-\min_{g \in
  G_s}\mu(g,\Psi^2_n(x_{V_n}))
\nonumber\\
 &\leq \min_{g \in
  G_s}\frac{1}{2}\rho(g,\Psi^1_n(x_{V_n}))-\min_{g \in
  G_s}h_b^{-1}\left(\rho(g,\Psi^2_n(x_{V_n})\right)
\nonumber\\
&= \frac{1}{2}\min_{g \in
  G_s}\rho(g,\Psi^1_n(x_{V_n}))-h_b^{-1}\left(\min_{g \in
    G_s}\rho(g,\Psi^2_n(x_{V_n}))\right).
\end{align}
Taking $\Psi^1_n$ to be the Peano-Hilbert scan, the results of
\cite{Lempel_Ziv86} imply that
\begin{equation}
\min_{g \in
G_s}\rho(g,\Psi_{PH}(x_{V_n})) \leq \min_{g \in G_s}\rho(g,\Psi_n(x_{V_n})) +
\epsilon_{n,s}
\end{equation}
for any finite-state scan $\Psi_n$, where $\epsilon_{n,s}$ satisfies $\lim_{s \to \infty}\limsup_{n \to \infty}\epsilon_{n,s} = 0$. Hence,
\begin{multline}
\min_{g \in G_s}\mu(g,\Psi_{PH}(x_{V_n}))-\min_{g \in
  G_s}\mu(g,\Psi(x_{V_n}))
\\
 \leq \frac{1}{2}\min_{g \in
  G_s}\rho(g,\Psi_{PH}(x_{V_n}))-h_b^{-1}\left(\min_{g \in
    G_s}\rho(g,\Psi_{PH}(x_{V_n}))-\epsilon_{n,s}\right).
\end{multline}
Taking the limits $\limsup_{n \rightarrow \infty}$ and then $s \rightarrow \infty$
implies the following proposition.
\begin{proposition}\label{prop. diff PH other Hamming loss}
Let $x$ be any individual image. Let $PH$ denote the Peano-Hilbert sequence of scans. Then, under the Hamming loss function, for any sequence of finite state scans $\Psi$ we have
\begin{equation} 
L_{PH}(x) \leq L_{\Psi}(x) + \frac{1}{2}\rho(x)-h_b^{-1}(\rho(x)),
\end{equation}
where $\rho(x)$ is the compressibility of the individual image $x$.
\end{proposition}
In other words, the specific
scandictor composed of the Peano-Hilbert
scan followed by the optimal predictor, adheres to the same asymptotic 
bounds (on predictability in terms of the compressibility) as the
\emph{best} finite-state scandictor. Figure \ref{fig. redundancy} plots the function
$\frac{1}{2}\rho-h_b^{-1}(\rho)$. The maximum
possible loss is $0.16$, similar to the bound given in Proposition \ref{prop. diff PH other}, yet this value is achieved only when the image's FS compressibility is around
$0.75$ bits/symbol. For images which are highly compressible, for example,
 when $\rho<0.1$ the resulting excess loss is smaller than $0.04$.

%% file: conclusion.tex
\section{Conclusion}\label{sec.conc}
In this paper, we formally defined finite set scandictability, and showed that there exists a universal algorithm which successfully competes with any finite set of scandictors when the random field is stationary. Moreover, the existence of a universal algorithm which achieves the scandictability of any spatially stationary random field was established. We then considered the scenario where non-optimal scanners are used, and derived a bound on the excess loss in that case, compared to optimal scandiction. 

It is clear that the scandiction problem is even more intricate than its prediction analogue. For instance, very basic results in the prediction scenario do not apply to the scandiction case in a straightforward way, and, in fact, are still open problems. To name a few, consider the case of universal scandiction of individual images, briefly discussed in Section \ref{subsec. universality for ind. images}. Although the question whether there exists a universal scandictor which competes successfully with any finite set of scandictors on any individual image was answered negatively in Section \ref{sec. negative result}, it is interesting to discover interesting sets of scandictors for which universal scandiction \emph{is possible}. The sequential prediction literature also includes an elegant result \cite{Fed_Mer_Gut92} on the asymptotic equivalence between finite state and Markov predictors. We conjecture that this equivalence does not hold in the multi-dimensional scenario for any individual image. Finally, the very basic problem of determining the optimal scandictor for a given random field $X$ with a \emph{known probability measure} $Q$, is still unsolved in the general case.  

It is also interesting to consider the problems of scanning and prediction, as well as filtering, in a noisy environment. These problems are intimately related to various problems in communications and image processing, such as filtering and denoising of images and video. As mentioned in Section \ref{sec. intro.}, these problems are the subject of \cite{Cohen_Weissman_Merhav_II07}.

%% file: appendix.tex
\appendix\section{Appendixes}
\subsection{Proof of Proposition \ref{prop. compete with a set of
    scan. block-wise}}\label{app. proof of prop. comp. with a set}
For the sake of simplicity, we suppress the dependence of $m(n)$ in $n$. Define
$W_i = \sum_{j=1}^{\lambda}{e^{-\eta L_{j,i}}}$. We have
\begin{eqnarray}
\log\frac{W_{(K+1)^2}}{W_0} &=& \log\sum_{j=1}^{\lambda}{e^{-\eta
L_{j,(K+1)^2}}} - \log\lambda
\nonumber\\
&\geq& \log \max_{j}e^{-\eta L_{j,(K+1)^2}}- \log\lambda
\nonumber\\
&=& -\eta \min_{j}L_{j,(K+1)^2}- \log\lambda
\nonumber\\
&=& -\eta L_{min}- \log\lambda. \label{eq. lower bound on log of ratio}
\end{eqnarray}
Moreover,
\begin{eqnarray}
\log\frac{W_{i+1}}{W_i} &=& \log
\frac{\sum_{j=1}^{\lambda}{e^{-\eta(L_{j,i}+L_j(x^i))}}}{\sum_{j'=1}^{\lambda}{e^{-\eta
L_{j',i}}}}
\nonumber\\
&=& \log
\sum_{j=1}^{\lambda}{P_i\left(j|\{L_{j,i}\}_{j=1}^{\lambda}\right)e^{-\eta
L_j(x^i)}}
\nonumber\\
&\leq& -\eta
\sum_{j=1}^{\lambda}{P_i\left(j|\{L_{j,i}\}_{j=1}^{\lambda}\right)L_j(x^i)}+\frac{m^4
l_{max}^2\eta^2}{8},
\end{eqnarray}
where the last inequality follows from the extension to Hoeffding's inequality
given in \cite{Mer_Orden_Serou_Weinb02} and the fact that $-\eta L_j(x^i)$ is
in the range $[-\eta m^2 l_{max},0]$. Thus,
\begin{eqnarray}
\log\frac{W_{(K+1)^2}}{W_0} &=&
\sum_{i=0}^{(K+1)^2-1}{\log\frac{W_{i+1}}{W_i}}
\nonumber\\
&\leq& -\eta \sum_{i=0}^{(K+1)^2-1}{\sum_{j=1}^{\lambda}
{P_i\left(j|\{L_{j,i}\}_{j=1}^{\lambda}\right)L_j(x^i)}}+\frac{m^4
l_{max}^2\eta^2(K+1)^2}{8}
\nonumber\\
&=& -\eta \bar{L}_{alg}+\frac{m^4 l_{max}^2\eta^2(K+1)^2}{8}. \label{eq. upper
bound on log of ratio}
\end{eqnarray}
Finally, from \eqref{eq. lower bound on log of ratio} and \eqref{eq. upper
bound on log of ratio}, we have
\begin{eqnarray}
\bar{L}_{alg} - L_{min} &\leq& \frac{\log\lambda}{\eta}+\frac{m^4
l_{max}^2\eta(K+1)^2}{8}
\nonumber\\
&\leq& \frac{\log\lambda}{\eta}+\frac{m^2 l_{max}^2\eta(n+m)^2}{8}. \label{eq.
both bounds on log of ratio}
\end{eqnarray}
The bound in \eqref{eq. regret in compete with a set of scan. block-wise}
easily follows after optimizing the right hand side of \eqref{eq. both
bounds on log of ratio} with respect to $\eta$.
%
%
\subsection{Proof of Proposition \ref{prop. compete with a set of
    scan. block-wise, a.s.}}\label{app. proof of prop. comp. with a
  set, a.s.}
Let $\delta(n)$ be some sequence satisfying $\delta(n) \to
0$ as $n \to \infty$. Define the sets
\begin{equation}
A_n=\left\{\omega:\frac{L_{alg}(x_{V_n})-L_{min}(x_{V_n})}{n^2}>\delta(n^2)\right\},
\end{equation}
where $(\Omega,P)$ is the probability space. We wish to show that
\begin{equation}
P\left(\limsup_{n \to \infty}A_n\right)=0,
\end{equation}
that is, $P(A_n \text{ i.o.})=0$. Let $(\Psi,F)_k$ be the scandictor chosen by the algorithm for the
$k+1$ block, $x^k$. Define
\begin{equation}
Z_k=L_{(\Psi,F)_k}(x^k)-E\left\{L_{(\Psi,F)_k}(x^k) | \{L_{j,k}\}_{j=1}^{\lambda}\right\},
\end{equation}
where the expectation is with respect to
$P_k\left(j|\{L_{j,k}\}_{j=1}^{\lambda}\right)$. Namely, the actual
randomization in $Z_k$ is in the \emph{choice} of $(\Psi,F)_k$. Thus,
$\{Z_k\}$ are clearly independent, and adhere to the following
Chernoff-like bound \cite[eq. 33]{Mer_Orden_Serou_Weinb02}
\begin{equation}
P\left(\sum_{k=1}^{(K+1)^2}{Z_k} \geq
  (K+1)^2\epsilon\right) \leq
\exp\left\{-\frac{2(K+1)^2\epsilon^2}{(m^2 l_{max})^2}\right\}
\end{equation}
for any $\epsilon>0$. Note that
\begin{equation}
\sum_{k=1}^{(K+1)^2}{Z_k} = L_{alg}(x_{V_n})-\bar{L}_{alg}(x_{V_n}),
\end{equation}
thus, together with eq. \eqref{eq. regret in compete with a set of
  scan. block-wise}, we have
\begin{multline}\label{eq. bound on tail prob.}
P\left(L_{alg}(x_{V_n}) - L_{min}(x_{V_n}) \geq
  (K+1)^2\epsilon + m(n+m)\sqrt{\log
\lambda}\frac{l_{max}}{\sqrt{2}}\right)
\\
\leq \exp\left\{-\frac{2(K+1)^2\epsilon^2}{(m^2 l_{max})^2}\right\}.
\end{multline}
Set
\begin{equation}
\delta(n)=\frac{(K+1)^2\epsilon + m(n+m)\sqrt{\log
\lambda}\frac{l_{max}}{\sqrt{2}}}{n^2}.
\end{equation}
Clearly $\delta(n) \to 0$ as $n \to \infty$ for any $m(n)=o(n)$ satisfying
$m(n) \to \infty$. For the summability of the r.h.s. of \eqref{eq. bound
  on tail prob.} we further require that
$m(n)=o\left(n^{1/3}\right)$. The proposition then follows directly by applying
the Borel-Cantelli lemma.
%
\subsection{Proof of Proposition \ref{prop. two identical scanners on B}}\label{app. proof of prop. two identical scanners on B}
We show, by induction on $w$, that the number of sites in $B$ for which the
context of size $w$ (in terms of sites in $B$) under the scan $\Psi$ is not
contained in the context of size $Kw$ under the scan $\Psi'$ is at most
$o(|B|)(K+1)^{w-1}$. This proves the proposition, as the cumulative loss of
$(\Psi',F^{Kw,opt})$ is no larger than $o(|B|)(K+1)^{w-1} l_{max}$ on these
sites, and is at least as small as that of $(\Psi,F^{w,opt})$ on all the rest
$|B|-o(|B|)(K+1)^{w-1}$ sites.

For $w=1$ this is indeed so, by our assumption on $\Psi$ and $\Psi'$ - i.e.,
\eqref{eq. cond. for identical scans}. We say that a site in $B$ satisfies the
context-condition with length $w-1$ if its context of size $i-1$, $1 < i \leq w$, under the scan
$\Psi$ is contained in its context of size $K(i-1)$ under the scan $\Psi'$.
Assume that the number of sites in $B$ which do not satisfy the
context-condition with length $w-1$ is at most $o(|B|)(K+1)^{w-2}$. We wish to
lower bound the number of sites in $B$ for which the context-condition with
length $w$ is satisfied. A sufficient condition is that the context-condition
with length $w-1$ is satisfied for both the site itself and its immediate past
under $\Psi$. If the context-condition with length $w-1$ is satisfied for a
site, its immediate past under $\Psi$ is contained in its past of length $K$
under $\Psi'$. Thus, if the context-condition of length $w-1$ is satisfied for
a given site, and for all $K$ preceding sites under $\Psi'$, then it is also
satisfied for length $w$. In other words, each site in $B$ which does not
satisfy the context-condition with length $w-1$ results in at most $K+1$ sites
(itself and $K$ more sites) which do not satisfy the context-condition with
length $w$. Hence, if our inductive assumption is satisfied for $w-1$, the
number of sites in $B$ which do not satisfy the context-condition with length
$w$ is at most $o(|B|)(K+1)^{w-2}(K+1)$, which completes the proof.
%
\subsection{Proof of Proposition \ref{cor. individual images 1}}\label{app. proof of cor. individual images 1}
The proof is a direct application of Propositions \ref{prop. compete with a set of scan. block-wise} and \ref{prop. two identical scanners on B}. For each $n$, define the scandictors set
\begin{eqnarray}
\tilde{\calF}_n&=&\Big\{(\Psi^1_n,F^{Kw,1}),(\Psi^1_n,F^{Kw,2}),\ldots,(\Psi^1_n,F^{Kw,|D|^{|A|^{Kw}}}),\ldots
\nonumber\\
&&\hspace{0.3cm}(\Psi^2_n,F^{Kw,1}),(\Psi^2_n,F^{Kw,2}),\ldots,(\Psi^2_n,F^{Kw,|D|^{|A|^{Kw}}}),\ldots
\nonumber\\
&&\hspace{0.3cm}\ldots
\nonumber\\
&&\hspace{0.3cm}(\Psi^\lambda_n,F^{Kw,1}),(\Psi^\lambda_n,F^{Kw,2}),\ldots,(\Psi^\lambda_n,F^{Kw,|D|^{|A|^{Kw}}})\Big\},
\end{eqnarray}
where $\{F^{Kw,i}\}_{i=1}^{|D|^{|A|^{Kw}}}$ is the set of all Markov predictors of order $Kw$.\footnote{Alternatively, one can use \emph{one} universal predictor which competes successfully with all the Markov predictors of that order.} Applying the results of Proposition \ref{prop. compete with a set of scan. block-wise} to $\{\tilde{\calF}_n\}$, we have, for any image $x$ and all $n$,
\begin{equation}\label{eq. prop. on finite set applied to individual}
EL_{(\hat{\Psi},\hat{F})_n}(x_{V_n}) - \min_{(\Psi,F) \in \tilde{\calF}_{m(n)}}
L_{(\Psi,F)}(x_{V_n}) \leq m(n)\left(n+m(n)\right)\sqrt{\log{\lambda
|D|^{|A|^{Kw}}}}\frac{l_{max}}{\sqrt{2}},
\end{equation}
where $\min_{(\Psi,F) \in \calF_{m(n)}} L_{(\Psi,F)}(x_{V_n})$ is the
cumulative loss of the best scandictor in $\calF_{m(n)}$ operating block-wise
on $x_{V_n}$. However, by Proposition \ref{prop. two identical scanners on B},
for any $1 \leq i \leq \lambda$, $x$ and $n$,
\begin{equation}
\min_{1 \leq j \leq |D|^{|A|^{Kw}}}EL_{(\Psi^i_{m(n)},F^{Kw,j})}(x_{V_n}) \leq
EL_{(\Psi^i_n,F^i)}(x_{V_n})+ o\left(m(n)^2\right)(K+1)^{w-1}
l_{max}\Floor[\frac{n}{m(n)}]^2.
\end{equation}
Note that
\begin{eqnarray}
\min_{(\Psi,F) \in \tilde{\calF}_{m(n)}} L_{(\Psi,F)}(x_{V_n}) &=& \min_{1 \leq
i \leq \lambda}\min_{1 \leq j \leq
|D|^{|A|^{Kw}}}EL_{(\Psi^i_{m(n)},F^{Kw,j})}(x_{V_n})
\nonumber\\
&\leq& \min_{1 \leq i \leq \lambda}\left\{ EL_{(\Psi^i_n,F^i)}(x_{V_n})+
o\left(m(n)^2\right)(K+1)^{w-1} l_{max}\Floor[\frac{n}{m(n)}]^2 \right\}
\nonumber\\
&=& \min_{(\Psi,F)\in \calF_n}EL_{(\Psi,F)}(x_{V_n}) +
o\left(m(n)^2\right)(K+1)^{w-1} l_{max}\Floor[\frac{n}{m(n)}]^2.
\nonumber\\
\end{eqnarray}
Thus, together with \eqref{eq. prop. on finite set applied to individual}, we have
\begin{multline}
EL_{(\hat{\Psi},\hat{F})_n}(x_{V_n})
- \min_{(\Psi,F)\in \calF_n}EL_{(\Psi,F)}(x_{V_n})
\\
\leq m(n)\left(n+m(n)\right)\sqrt{\log{\lambda
|D|^{|A|^{Kw}}}}\frac{l_{max}}{\sqrt{2}}+ o\left(m(n)^2\right)(K+1)^{w-1}
l_{max}\Floor[\frac{n}{m(n)}]^2,
\end{multline}
which completes the proof since $|D|,|A|,K$ and $w$ are finite.
%
\subsection{Proof of Proposition \ref{prop. abs diff entropy and loss with scanning, ind. sequence}}\label{app. proof of prop. abs diff entropy and loss with scanning, ind. sequence}
Similar to the proof of Proposition \ref{prop. abs diff entropy and loss}, we have,
\begin{eqnarray}
&&\hspace{-1cm} \left|\alpha_l \hat{H}_{\Psi_{B}}^{k+1}(X|X^k) +\beta_l-\frac{1}{|B|}L_{(\Psi_{B},F^{k,opt})}(x_B)\right|
\nonumber\\
&&\hspace{-0.5cm}=\left|\alpha_l \hat{H}_{\Psi_{B}}^{k+1}(X|X^k) +\beta_l - \frac{1}{|B|}\left(\sum_{t=1}^{k}{l(x_t,F^{k,opt}(x_1^{t-1}))} + \sum_{t=k+1}^{|B|}{l(x_t,F^{k,opt}(x_{t-k}^{t-1}))}\right) \right|
\nonumber\\
&&\hspace{-0.5cm}=\Bigg|\alpha_l \hat{H}_{\Psi_{B}}^{k+1}(X|X^k) +\beta_l -
 \nonumber\\
&&\hspace{1cm}
 \frac{1}{|B|}\sum_{t=1}^{k}{l(x_t,F^{k,opt}(x_1^{t-1}))} - \left(1-\frac{k}{|B|}\right)\frac{1}{|B|-k}\sum_{t=k+1}^{|B|}{l(x_t,F^{k,opt}(x_{t-k}^{t-1}))} \Bigg|
\nonumber\\
&&\hspace{-0.5cm}\leq \left|\alpha_l \hat{H}_{\Psi_{B}}^{k+1}(X|X^k) +\beta_l - \frac{1}{|B|-k}\sum_{t=k+1}^{|B|}{l(x_t,F^{k,opt}(x_{t-k}^{t-1}))} \right| + \frac{k l_{max}}{|B|}. 
\end{eqnarray}
Since the order of the predictor is fixed, we can use the definition of $\hat{P}_{\Psi_{B}}^{k+1}(s)$ ans sum over $s \in \{0,1\}^{k+1}$ instead of $t$. Thus,
\begin{eqnarray}
&&\hspace{-1cm} \left|\alpha_l \hat{H}_{\Psi_{B}}^{k+1}(X|X^k) +\beta_l-\frac{1}{|B|}L_{(\Psi_{B},F^{k,opt})}(x_B)\right|
\nonumber\\
&&\hspace{-0.5cm} \leq \left|\alpha_l \hat{H}_{\Psi_{B}}^{k+1}(X|X^k) +\beta_l - \sum_{s \in \{0,1\}^{k+1}}{\hat{P}_{\Psi_{B}}^{k+1}(s) l(s_{k+1},F^{k,opt}(s_1^k))} \right| + \frac{k l_{max}}{|B|} 
\nonumber\\
&&\hspace{-0.5cm}=\left|\sum_{s' \in \{0,1\}^k}{\hat{P}_{\Psi_{B}}^{k+1}(s') \sum_{x \in \{0,1\}}{\hat{P}_{\Psi_{B}}^{k+1}(x|s') \left(-\alpha_l \log \hat{P}_{\Psi_{B}}^{k+1}(x|s')  + \beta_l -  l(x,F^{k,opt}(s'))\right)}} \right|
\nonumber\\
&&\hspace{1cm} + \frac{k l_{max}}{|B|}
 \nonumber\\
&&\hspace{-0.5cm}=\left|\sum_{s' \in \{0,1\}^k}{\hat{P}_{\Psi_{B}}^{k+1}(s')  \left(\alpha_l h_b( \hat{P}_{\Psi_{B}}^{k+1}(\cdot|s'))  + \beta_l -  \phi_l(\hat{P}_{\Psi_{B}}^{k+1}(\cdot|s'))\right)} \right| + \frac{k l_{max}}{|B|}
 \nonumber\\
&&\hspace{-0.5cm} \leq \sum_{s' \in \{0,1\}^k}{\hat{P}_{\Psi_{B}}^{k+1}(s')  \max_p \left| \alpha_l h_b(p)  + \beta_l -  \phi_l(p)\right|} + \frac{k l_{max}}{|B|}
\nonumber\\
&&\hspace{-0.5cm} = \epsilon_l + \frac{k l_{max}}{|B|}.
\end{eqnarray}